\documentclass{article}
\usepackage{ourmacros}

\title{Tensor Computation of Euler Characteristic Functions and Transforms} 
\date{}

\author{Jessi Cisewski-Kehe\thanks{Department of Statistics, University of Wisconsin-Madison (\texttt{jjkehe@wisc.edu}).}
  \and Brittany Terese Fasy\thanks{School of Computing, Montana State University (\texttt{brittany.fasy@montana.edu}), (\texttt{alex.mccleary@gmail.com}), (\texttt{eli.quist@student.montana.edu}).}
  \and Alexander McCleary\footnotemark[2]
  \and Eli Quist\footnotemark[2]}

\begin{document}

\maketitle

\begin{abstract}
The weighted Euler characteristic transform (WECT) and 
Euler characteristic function~(ECF)
have proven to be useful tools in a variety of applications.
However, current methods for computing these functions are either not optimized
for GPU computation or do not scale to higher-dimensional settings.
In this work, we present a tensor-based framework for computing such topological
descriptors which is highly optimized for GPU
architectures and works in full generality across 
simplicial and cubical complexes of arbitrary dimension.
Experimentally, the framework demonstrates significant speedups over existing
methods when computing the WECT and ECF across a variety of two- and
three-dimensional datasets.
Computation of these transforms is implemented in a publicly available Python
package called \texttt{pyECT}.\footnote{
    \url{https://github.com/compTAG/pyECT}
}

\end{abstract}

\section{Introduction} \label{sec:intro}

The field of topological data analysis (TDA) utilizes invariants from topology
to study the shape of data.
Among the invariants employed are homology and the original
topological invariant, the Euler characteristic~\cite{euler1758elementa}.
In practice, a single invariant is not typically robust enough to capture the
shape of data.
For this reason, filtrations (i.e., nested sequences of shapes) are often
constructed from data and topological invariants are then applied.
This allows for significantly more information to be retained, leading to
constructions such as persistence diagrams \cite{edel2002topological}
and Euler characteristic
functions \cite{Serra1982}.

Topological transforms \cite{Turner2014PHT} further extend the idea of applying
topological invariants to filtrations by constructing a parameterized family of
filtrations and then applying topological invariants.
The starting point for topological transforms is typically a geometric simplicial or cubical
complex~$K$ in $\R^{n}$.
For each direction vector $\vect{s} \in \sph^{n-1}$, a directional filtration of~$K$
is constructed and then a topological invariant is applied to that
filtration.
Applying the Euler characteristic~(EC) to a single directional filtration
results in an Euler characteristic function (ECF),
while applying it to each directional filtration in~$\sph^{n-1}$
results in the Euler characteristic transform~(ECT) \cite{Turner2014PHT,munch2025invitation}.
The ECT is a \emph{complete invariant}: a geometric simplicial complex $K$
in $\R^{n}$ can be reconstructed solely from
its~ECT~\cite{Ghrist2018,schapira1995tomography}.
In other words, no information is lost when passing from the simplicial
complex~$K$ to its ECT.

Useful variants of the ECF and ECT are the weighted Euler characteristic
function~(WECF) and the weighted Euler characteristic transform (WECT)
for weighted complexes~\cite{Jiang2020TheWE}.
Much like the ECT, the WECT is a complete invariant for weighted geometric
complexes~\cite{Jiang2020TheWE}.
Moreover,
the WECT has been employed in a broad range of applications,
including: in oncology for predicting survival rates of
patients with glioblastoma multiforme brain
tumors~\cite{crawford2020predicting}, in
plant biology for classifying seed phenotypes
\cite{amezquita2022measuring}, and in image classification
tasks~\cite{betthauser2018topological, cisewski2023weighted, Jiang2020TheWE}.
Numerous software packages already exist for computing
the WECT and its variants, including those presented in
\cite{amezquita2022measuring, crawford2020predicting, laky2024,
lebovici2024efficientcomputationtopologicalintegral, roell2023differentiable,
saxena2025scalable}.
While many of these packages provide useful tools, they are either not optimized
for GPU architectures or are limited in the types (or dimension) of
complexes they can process.

In this work, we provide a fully tensor-based framework for computing WECFs and
WECTs.
Our implementation applies to arbitrary geometric simplicial
and cubical complexes in Euclidean spaces.
The tensor operations we use are highly optimized for fast computation on
modern GPU hardware, providing significant speedups over existing~implementations.

\section{Background}

In this section, we provide necessary foundations in
computational topology and tensor computation to develop the main results.
For simplicity, we restrict our attention to (abstract) simplicial complexes;
however, our results and code also work for cubical complexes.

For a simplicial complex $K$ and for any
$i \in \{0,1,\ldots, \dim(K)\}$, we denote the set of~$i$-simplices of $K$
by~$K_i$.
We use the notation $\sub(K)$ for the set of subcomplexes of $K$.
Furthermore, we define~$k = |K|$ and~$k_i = |K_i|$.

\begin{definition}[Lower-Star Filtrations]\label{def:lower_star_filtration}
    Let $K$ be a simplicial complex with vertex set~$K_0$.
    A \emph{vertex filter} on $K$ is a function~${f \colon K_0 \to \R}$.
    For a vertex filter $f\colon K_0 \to \R$,
    the \emph{lower-star filtration} induced by $f$ is the
    function $F\colon \R \to \sub(K)$,
    defined as
    \begin{equation}
        F(t) := \{ \sigma \in K \mid \max_{v \in \sigma} f(v) \le t \}.
    \end{equation}
\end{definition}

The main examples of lower-star filtrations used in this paper are directional
filtrations.
These are defined for any (abstract) simplicial complex $K$ with a function
$\ell\colon K_0 \to \R^n$.
In practice, simplicial complexes are often naturally embedded in $\R^n$ and,
in this setting, the function $\ell$ simply maps each vertex to its coordinates.

\begin{definition}[Vertex Height Functions and Directional Filtrations]
    Let $K$ be a simplicial complex with a map~$\ell \colon K_0 \to \R^n$ and
    let~${\vect{s} \in \sph^{n-1}}$ be a
    direction vector.
    The \emph{vertex height function} in direction $\vect{s}$ is the
    map~${\heightfcn{\vect s} \colon K_0 \to \R}$ defined as
    $
        \heightfcn{\vect s}(v) := \ell(v) \cdot \vect{s}.
    $
    The \emph{directional filtration} of $K$ in direction~$\vect{s}$, denoted
    as $H_{\vect{s}}$, is the lower-star filtration of $K$ induced by
    $h_{\vect s}$.
\end{definition}

\subsection{Weighted Euler Characteristic Functions and Transforms}

Here, we introduce weighted Euler characteristic functions and transforms.
For a more thorough review, see the survey paper \cite{munch2025invitation}.

\begin{definition}[Weighted Simplicial Complexes]
    A \emph{weighted simplicial complex} is a pair~$(K, \omega)$, where $K$ is
    a simplicial complex and $\omega \colon K \to \R$ is a function
    called a \emph{weight function}.
\end{definition}

Weighted simplicial complexes are often used to model various types of data.
For example, triangular meshes are widely used to model
three-dimensional surfaces,
with weight functions capturing scalar fields on the surface.
Another example comes from image data, where grayscale images are converted
into weighted simplicial complexes via the \emph{Freudenthal triangulation}
\cite{cisewski2023weighted, Jiang2020TheWE}
with weights derived from the pixel intensities.

\begin{definition}[Weighted Euler Characteristic]
    The \emph{weighted Euler characteristic~(WEC)} of a weighted simplicial
    complex~$(K, \omega)$ is
    \[
        \chi(K, \omega) := \sum_{\sigma \in K} (-1)^{\dim{\sigma}}
        \omega(\sigma).
    \]
\end{definition}
Note that the Euler characteristic is a special case of the WEC,
where the weight function is equal to the constant function
$\mathbf{1} \colon K \to \R$.
\journal{
\begin{remark}\label{rem:ec-wec}
    The Euler characteristic is a special case of the WEC.
    For a simplicial complex~$K$, its Euler characteristic is the WEC of
    $(K, \mathbf{1})$, where $\mathbf{1} \colon K \to \R$ is the constant
    function
    that assigns value one to each simplex.
\end{remark}
}
While the WEC is a remarkably useful invariant given its simplicity, there is
limited information that a single number can capture about a simplicial complex.
To incorporate more information, we apply the WEC to a
filtration:

\begin{definition}[Weighted Euler Characteristic Function]
    Let $\K = (K, \omega)$ be a weighted simplicial complex
    and let $f \colon K_0 \to \R$ be a vertex filter of $K$ with
    lower-star filtration $F$.
    The \emph{weighted Euler characteristic function}\footnote{
        Weighted Euler characteristic functions are also known as weighted
        Euler characteristic curves.
        Due to their general lack of continuity, we use the term function
        rather than curve.
        }
    of $\K$ induced by $f$ is
    the function
    $\wecf_{f} \colon \R \to \R$
    defined as
    \begin{equation}
        \wecf_{f}(t) := \chi \big( F(t), \omega|_{F(t)} \big). \label{eq:ecf}
    \end{equation}
\end{definition}

\begin{definition}[Directional WECF]
    Let $\K = (K, \omega)$ be a weighted simplicial complex with a function~\mbox{$\ell\colon K_0 \to \R^n$}
    and let~${\vect s \in \sph^{n-1}}$ be a direction vector.
    The \emph{directional WECF} of $\K$ in direction~$\vect s$ is the WECF of
    $\K$ induced by the height function $h_\vect s$.
\end{definition}

\begin{remark}\label{rem:which-ecf}
    In image analysis tasks (and their three-dimensional analogs),
    there are two common (W)ECFs considered.
    The first (implemented in
    \cite{laky2024, wang2023gpu,wang2023gpu_conference}) uses the pixel
    intensities as the vertex filter.
    We refer to the resulting ECF simply as the ECF of an image.
    The second, which we refer to as the directional WECF of an image, uses the
    pixel intensities as weights
    and the height function $h_\vect s$ for a given direction vector
    $\vect{s} \in \sph^1$ as the vertex filter.
    This method is implemented in
    \cite{amezquita2022measuring, tang2022topological, lebovici2024efficientcomputationtopologicalintegral}.
\end{remark}

\begin{figure}[htbp]
    \centering
    \begin{subfigure}[t]{0.2\textwidth}
        \centering
        \includegraphics[height=.95in]{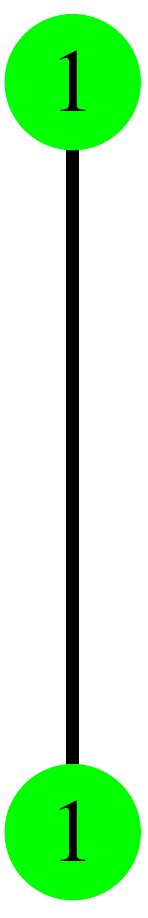}
        \subcaption{Time=0}
        \label{fig:ecf_example-simp0}
    \end{subfigure}%
    ~
    \begin{subfigure}[t]{0.3\textwidth}
        \centering
        \includegraphics[height=.95in]{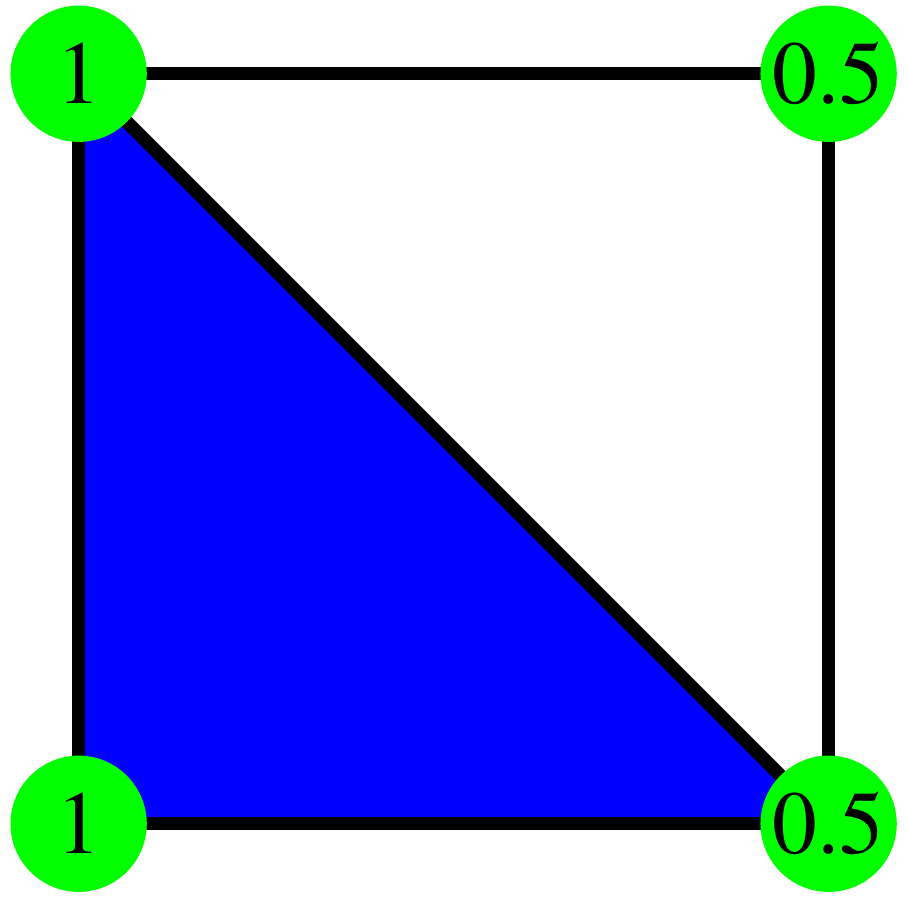}
        \subcaption{Time=1}
        \label{fig:ecf_example-simp1}
    \end{subfigure}
    ~
    \begin{subfigure}[t]{0.35\textwidth}
        \centering
        \includegraphics[height=.95in]{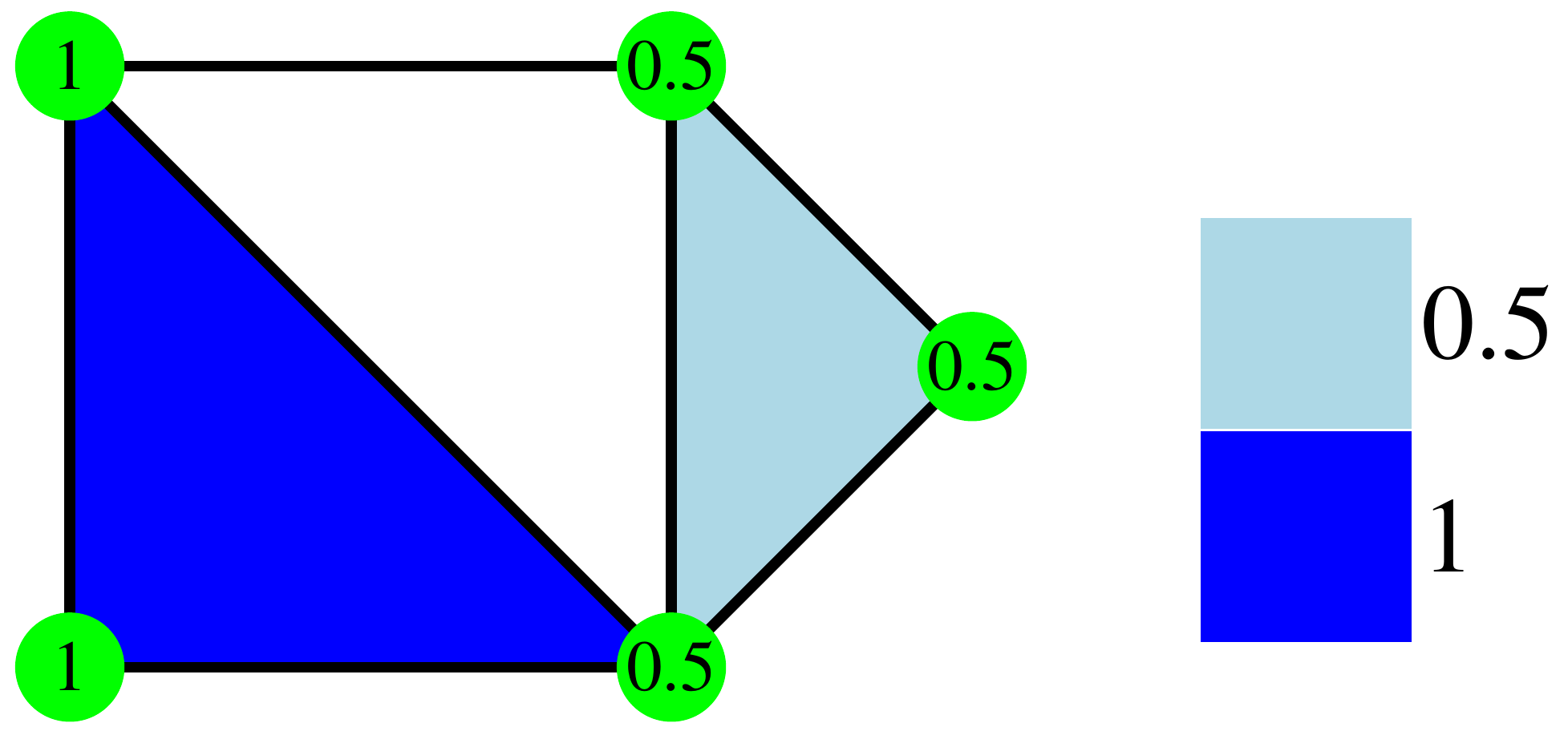}
        \subcaption{Time=2}
        \label{fig:ecf_example-ecf_simplex_full}
    \end{subfigure}
    \\
    \begin{subfigure}[t]{0.45\textwidth}
        \centering
        \includegraphics[height=1.5in]{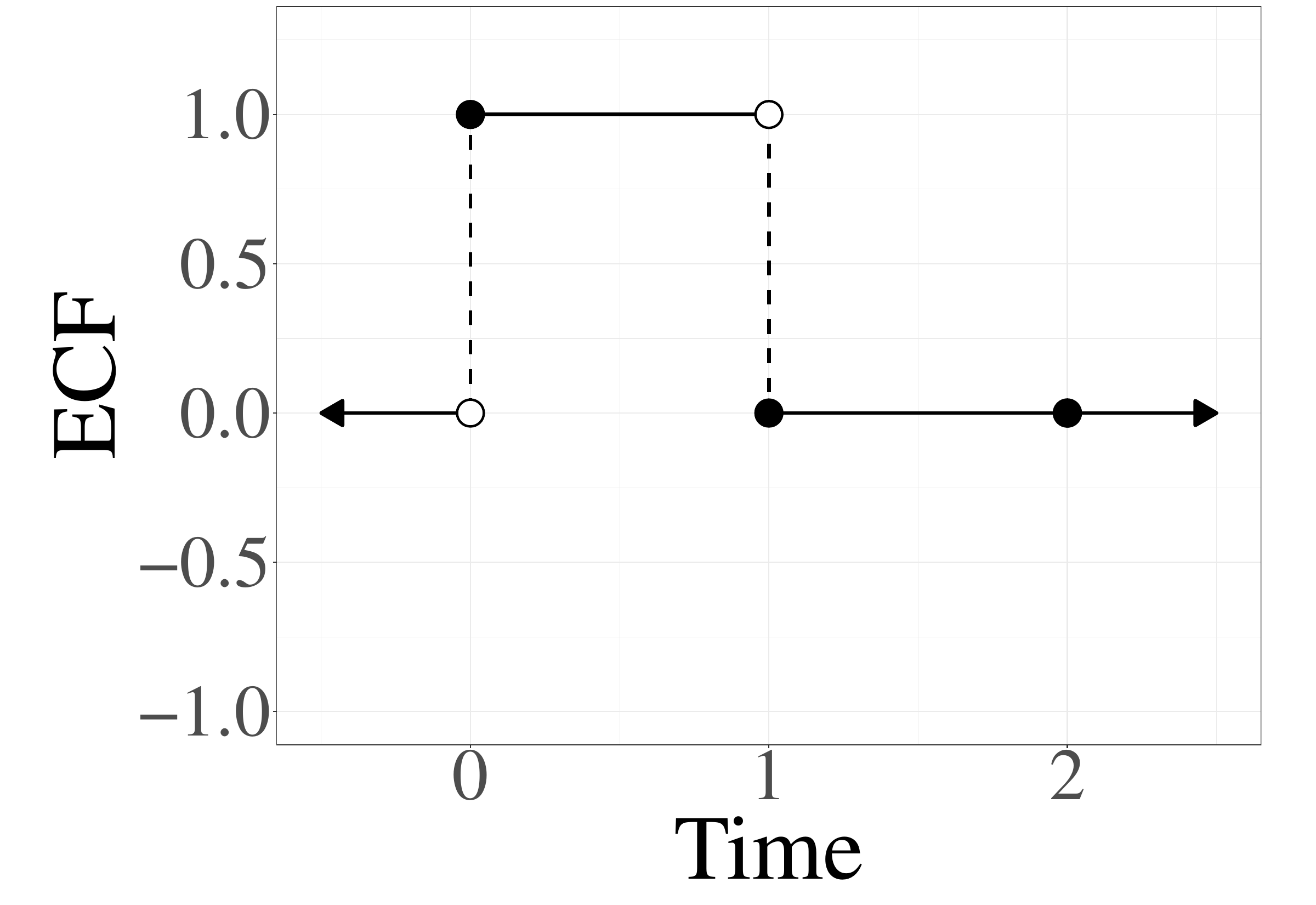}
        \subcaption{ECF in direction (1,0)}
        \label{fig:ecf_example-ecf_ex}
    \end{subfigure}
    ~
    \begin{subfigure}[t]{0.45\textwidth}
        \centering
        \includegraphics[height=1.5in]{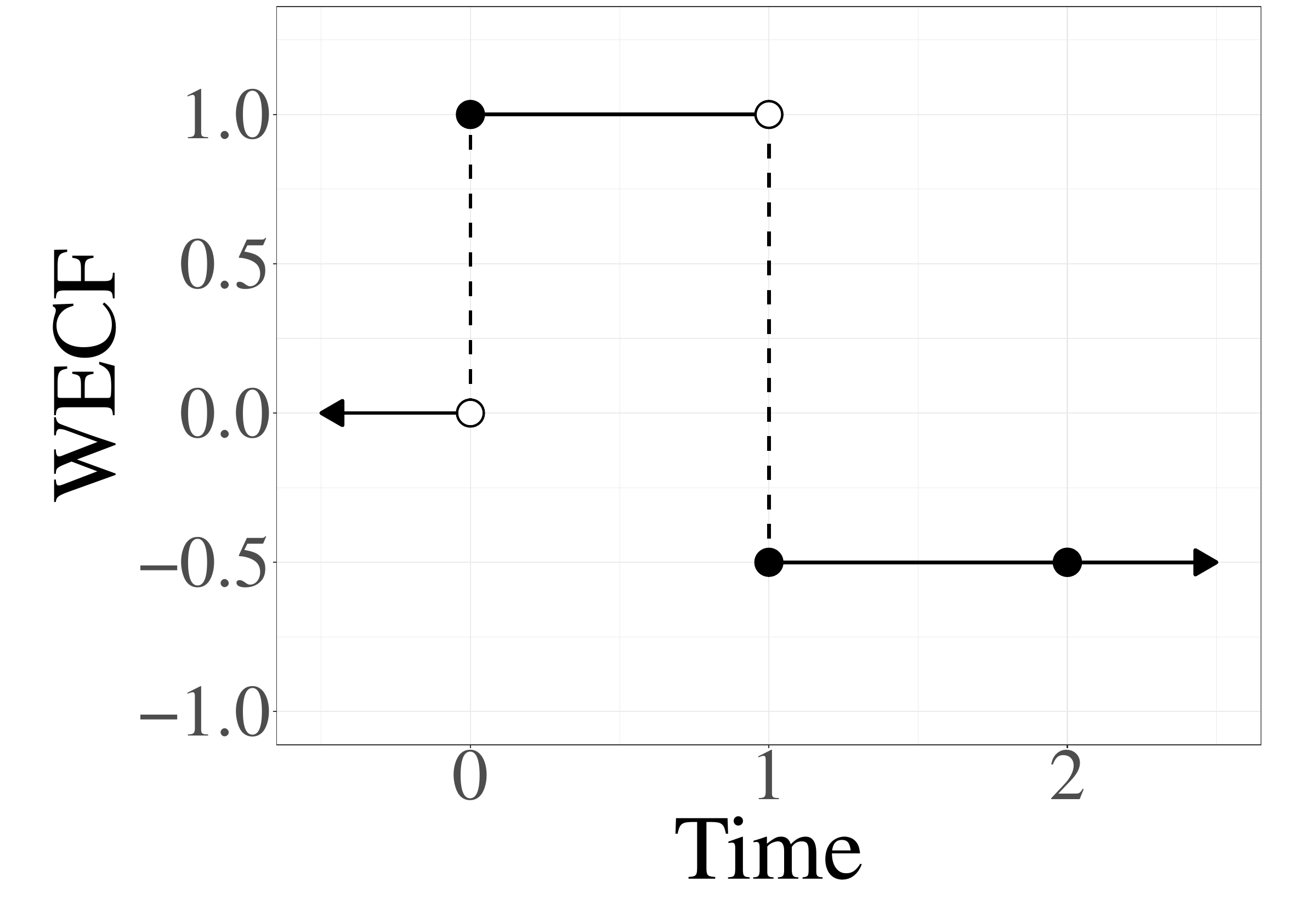}
        \subcaption{WECF in direction (1,0)}
        \label{fig:ecf_example-wecf_ex}
    \end{subfigure}
    \caption{
        Filtration of a simplicial complex
        (\subref{fig:ecf_example-simp0})--(\subref{fig:ecf_example-ecf_simplex_full})
        along direction (1,0), with vertex weights shown in green discs.
        The weight assigned to a simplex is the maximum weight of
        its vertices.
        The corresponding ECF and WECF are displayed in
        (\subref{fig:ecf_example-ecf_ex}) and (\subref{fig:ecf_example-wecf_ex})
        respectively.
        } \label{fig:ecf_example}
\end{figure}

\begin{example}
    An illustration of the directional ECF and directional WECF is presented
    in \figref{ecf_example}.
    The full simplicial complex, including weighted vertices,
    is displayed in \subfigref{ecf_example}{ecf_simplex_full}.
    The filtration is along direction~$(1,0)$,
    and the corresponding ECF and WECF are included in
    \subfigref{ecf_example}{ecf_ex}
    and \subfigref{ecf_example}{wecf_ex}, respectively.
    The ECF and WECF are the same until time 1 when two new vertices
    are introduced with weights of 0.5.
\end{example}

Typically, there is no canonical direction vector to choose when computing
directional WECFs.
Hence, it is common to choose \emph{all} directions,
resulting in the weighted Euler characteristic transform (WECT).

\begin{definition}[Weighted Euler Characteristic Transform]\label{def:wect}
    Let $\K = (K, \omega)$ be a weighted simplicial complex with a map
    $\ell\colon K_0 \to \R^{n}$.
    The weighted Euler characteristic transform of~$\K$ is the function
    $$\wect_{\K} \colon \sph^{n-1} \times \R \to \R$$
    defined as
    \begin{equation}
        \wect_{\K}(\vect{s}, t) := \wecf_{h_{\vect s}} (t)
        = \chi \big( H_{\vect{s}} (t), \omega \big).
    \end{equation}
\end{definition}

%
\subsection{Existing Frameworks for Computation of the ECF and (W)ECT}
\label{sec:existing-frameworks}

Due to its simplicity and ease of computation, the WECT and its variants are
widely used in applied settings (resulting in many available software packages).
%
\demeter \cite{amezquita2022measuring} is a Python package that
computes the directional ECF and ECT of axis-aligned cubical complexes.
Note that it incorporates an optimization of bucket sorting
the filtration values of each vertex.
However, this package does not work for generic simplicial complexes,
and only computes the special case of the unweighted ECT.
Next, the \sinatra package \cite{tang2022topological} is an R package that
computes the directional ECF and ECT through an algorithm similar to \demeter.
However, unlike \demeter, \sinatra works for arbitrary, unweighted
simplicial complexes.
Note that as \demeter and \sinatra are not optimized for speed, they are
not included in the experimental comparison.
Third, \eucalc%
\footnote{
    \url{https://github.com/HugoPasse/Eucalc}
} \cite{lebovici2024efficientcomputationtopologicalintegral}
is a highly optimized package that uses a preprocessing step to
compute critical points of the directional ECF.
This allows for efficient computation of the exact directional ECF
(as opposed to a vectorization of it).
The package only accepts axis-aligned cubical complexes as input,
but does generalize to weighted complexes.
Finally, the \dect package%
\footnote{
    \url{https://github.com/aidos-lab/dect}
}
\cite{roell2023differentiable} is a Python package
designed to compute the differentiable ECT (DECT) of simplicial complexes.
While the DECT is a slightly different variant, the package also includes
an implementation for computing the ECT of simplicial or cubical complexes.
Although it does not allow for weighted complexes, this package is designed for
GPU architectures, making it an interesting point of comparison.
Furthermore, \cite{saxena2025scalable} provides further optimizations for
computing the DECT on CUDA GPUs, but is consequently a far
less general framework.

The \fasttopology framework%
\footnote{
    \url{https://github.com/zavalab/ML/tree/master/FastTopology}
} \cite{laky2024} is a highly optimized
implementation for computing the ECF of lower-star filtrations of two- and
three-dimensional axis-aligned cubical complexes.
While it is designed to be highly parallelizable on CPU architectures,
it does not directly support vectorized computation (e.g., on a GPU)
or the WECF.
\gpuecc \cite{wang2023gpu, wang2023gpu_conference} is a GPU-based implementation
of the ECF.
Unlike the other existing packages, it is designed for GPU architectures.
However, it is highly optimized for image data, so it only works for
axis-aligned cubical complexes and does not generalize to weighted complexes.
Additionally, it does not compute the directional ECF or ECT,
and currently is only available as an implementation on Windows systems
(which is unavailable in our computation environment).

The framework presented in this paper, \pyect, admits several advantages
over these existing packages for computing the WECT and ECF.
First, unlike all existing frameworks (except \sinatra), \pyect works in
full generality across arbitrary weighted simplicial and cubical complexes;
see \tabref{input-comparison}.
Second, \pyect provides methods for computing both the WECFs of lower-star
filtrations and WECTs of weighted geometric complexes.
This generality is not available in any of these existing packages.
Third, the tensor operations used in \pyect are highly optimized
for modern GPU hardware, resulting in significant speedups
over these existing packages.
The \gpuecc package admits this advantage as well, but is far less general as it
only accepts axis-aligned cubical complexes and does not compute the directional
WECF or WECT.
Finally, the tensor-based approach used in \pyect yields simple code
(our core WECT algorithm is around $25$ lines of code) that is easily readable,
maintable, and customizable.

One minor disadvantage of \pyect is that it does not compute exact
EC values at all heights of the directional ECF
(which is available in \eucalc).
While computing exact values retains more information and can be an advantage
for reconstruction tasks, it leads to difficulties in
vectorizing the (W)ECT.
However, vectorizations of the WECT are in fact more desirable for many
real-world applications, especially when used as input to distance-based
methods or machine learning pipelines (such as those in
\cite{Jiang2020TheWE, Turner2014PHT, munch2025invitation, amezquita2022measuring}).

\begin{table}[ht]
    \centering
    \caption{
        Comparison of supported input types for ECT packages.
    }
    \begin{tabular}{l llll}
        \toprule
        Package
            & \shortstack[l]{Simplicial \\ Complexes}
            & \shortstack[l]{Cubical \\ Complexes}
            & \shortstack[l]{Axis-Aligned \\ Cubical Complexes}
            \\
        \midrule
        \pyect       & \yes & \yes & \yes  \\
        \eucalc       & \no & \no & \yes \\
        \fasttopology & \no & \no & \yes \\
        \demeter         & \no & \no & \yes\\
        \sinatra        & \yes & \yes & \yes\\
        \bottomrule
    \end{tabular}
    \label{tab:input-comparison}
\end{table}

%
\subsection{Tensor Operations}\label{ss:tensorops}

Our algorithm almost exclusively uses tensor operations.
These operations have highly
optimized implementations in PyTorch~\cite{pytorch1,ansel2024pytorch} that
run in parallel on both CPUs and GPUs.
Here, we review tensors and the tensor operations used in our algorithm
(see \appendref{tensor-examples} for examples).

For any positive integer $m$, let $[[m]] := \{0, 1, \ldots, m-1\}$.
For $\{m_i \}_{i=0}^{d-1} \subset \Z_{+}$, a
$d$-dimensional tensor of shape~$(m_0, m_1, \ldots, m_{d-1})$ is a function
\[
    T \colon [[m_0]] \times [[m_1]] \times \cdots \times [[m_{d-1}]] \to \R.
\]
We exclusively use one-dimensional tensors (vectors), two-dimensional
tensors (matrices), and three-dimensional tensors.
A specific dimension of a tensor is called an \emph{axis}.
We now introduce the tensor operations utilized in our algorithm.

\begin{enumerate}
    \item Zeros: Given a shape $L$, $\zeros(L)$ is the
        tensor with shape~$L$, where every value is~$0$.
    \item Element-wise operations: Any function $f \colon \R \to \R$ can be
        applied to
        tensors element-wise by post-composition. That is, if $T$ is a tensor,
        then
        $f(T)$ is the tensor with the same shape as $T$ and values
        $$
            f(T)[i_0, i_1, \ldots, i_{d-1}] =
            f \big( T[i_0, i_1, \ldots, i_{d-1}]
            \big).
        $$
    \item Matrix multiplication: If $T$ and $S$ are tensors with shape
        $(m_0, m_1)$ and $(m_1, m_2)$ respectively, we use the notation $T * S$
        to denote their matrix multiplication.
    \item Cumulative sum: Let $T$ be a tensor of shape $(m_0, m_1)$.
        Then, $\cumsum(T)$ is the tensor with shape $(m_0, m_1)$ and values
        \[
            \cumsum(T)[i,j] = \sum_{\substack{k \in [[m_1]] \\ k \leq j}} T[i,k].
        \]
        In this work, we use the cumulative sum on two-dimensional
        tensors along the second axis.
        However, the cumulative sum can be defined over
        any axis of a~tensor.
    \item Reduce maximum: Let $T$ be a tensor of shape
    $(m_0, m_1, \ldots, m_{d-1})$
        and let $p \in [[d]]$.
        Then, $\rmax (T, p)$ is the tensor with shape
        $
            (m_0, \ldots, m_{p-1}, m_{p+1}, \ldots, m_{d-1})
        $
        and values
        \[
            \rmax(T,p)[i_0, \ldots, i_{p-1}, i_{p+1}, \ldots i_{d-1}] =
            \max_{j \in [[m_p]]}T[i_0, \ldots, i_{p-1}, j, i_{p+1},
            \ldots i_{d-1}].
        \]
    \item Advanced indexing: Let $T$ be a tensor with shape $(m_2, m_3)$, and
        let $I$ be a tensor with shape $(m_0, m_1)$ and values in $[[m_2]]$.
        The tensor $T[I]$ has shape~$(m_0,m_1,m_3)$ and~values
        \[
            T[I][i,j,k] = T \big[ I[i,j], k \big].
        \]
    \item Scatter add: Let $S$ be a tensor with shape $(m_0,m_1)$, let $I$ be a
        tensor with shape $(m_0,m_2)$ and values in $[[m_1]]$, and let $T$
        be a tensor with shape~$(m_2)$.
        The function~$\scadd(I,T)$ updates $S$ in-place by adding
        to it the difference tensor~$D$ with shape $(m_0,m_1)$ and values
        \begin{equation}\label{eq:difftensor}
            D[i,j] = \sum_{\substack{k \in [[m_2]] \\ I[i, k] = j}} T[k].
        \end{equation}
        That is, for a tensor $S$, we have
        $
            \scadd(I,T)(S) = S + D,
        $
        where $+$ is taken element-wise and $S$ is updated in place.
        In practice, the tensor $D$ is not explicitly built.
\end{enumerate}

\section{Computation of WECFs}\label{sec: computation}
In this section, we present a new algorithm for computing the WECFs of a
weighted simplicial complex with respect to a finite set of vertex filters.
Computation of the WECT of a weighted geometric simplicial complex is a
special case of \algref{WECT}, and corresponds to the case where the vertex
filters are height functions; see Example~\ref{ex:wect}.

We fix a weighted simplicial complex $\K= (K, \omega)$ and a set of vertex
filters $\{f_0, f_1, \ldots, f_{m-1}\}$ for the entirety of this section.
To simplify notation, we extend each $f_p \colon K_0 \to \R$ to $K$ by
defining $f_p^{\max} \colon K \to \R$ as
\[
    f_p^{\max} (\sigma) = \max_{v \in \sigma} f_p(v).
\]
We fix enumerations $K_0 = \{v_0, v_1, \ldots, v_{k_0 - 1}\}$
and~$K_i = \{\sigma_0^i, \sigma_1^i, \ldots, \sigma_{k_i - 1}^i \}$ for each
$i \geq 1$.

Our algorithm uses the following inputs
\begin{itemize}
    \item $\complex$: A list of tuples:
    \begin{itemize}
        \item $\complex[0] = (\fvals, \vweights)$
        \begin{itemize}
            \item $\fvals$ is a tensor with shape $(k_0, m)$ and values
            $
                \fvals[a,p] = f_p(v_a).
            $
            \item $\vweights$ is a tensor with shape $(k_0)$ and values
            $
                \vweights[a] = \omega(v_a).
            $
        \end{itemize}

        \item for $i \geq 1$: $\complex[i] =
        (i\texttt{-SimplexVertices}, i\texttt{-SimplexWeights})$
        \begin{itemize}
            \item $i\texttt{-SimplexVertices}$ is a tensor with shape
                $(k_i, i+1)$ and values in~$[[k_0]]$.
                The rows of this tensor correspond to the
                vertex sets of each $i$-simplex.
            \item $i\texttt{-SimplexWeights}$ is a tensor with shape $(k_i)$
                and values
            $
                i\texttt{-SimplexWeights}[b] = \omega(\sigma_b^i).
            $
        \end{itemize}
    \end{itemize}
    \item $\numvals$: The number of values to sample the filter functions
    over.
\end{itemize}

In order to discretize the filtration height parameter,
define $\maxheight$ as
\[
    \maxheight = \max_{\substack{p \in [[m]] \\ v \in K_0}} |f_p(v)|.
\]
We discretize $[-\maxheight, \maxheight]$ by evenly sampling $\numvals$ points.
That is, we define the~map
$$
    \beta \colon [[\numvals]] \to [-\maxheight, \maxheight]
$$
by
\[
    \beta (q) = q \frac{2 * \maxheight} {\numvals - 1} - \maxheight.
\]
The function $\beta$ has a left inverse
$
    \alpha \colon [-\maxheight, \maxheight] \to [[\numvals]]
$
defined as
\begin{equation} \label{eq:left-adjoint}
    \alpha(t) = \bigg\lceil \frac{(\numvals - 1)
    (\maxheight + t)}{2 * \maxheight} \bigg\rceil.
\end{equation}
The maps $\alpha$ and $\beta$ are order-preserving and satisfy
\begin{equation} \label{eq:galois}
    \alpha(t) \leq q \iff t \leq \beta(q)
\end{equation}
for all $t \in [-\maxheight, \maxheight]$ and
$q \in [[\numvals]]$.\footnote{
    In other words, $\alpha$ and $\beta$ form a Galois connection.
}

\begin{algorithm}[htbp]
    \caption{\texttt{ComputeWECFs}$(\complex, \numvals)$}
    \label{alg:WECT}
    \begin{algorithmic}[1]
        \REQUIRE $\complex$,
                 $\numvals$
        \ENSURE $\wecfs$: a tensor of shape $(m, \numvals)$ and values
        $\wecfs[p, q] = \wecf_{f_p}( \beta(q))$.

        \vspace{1ex}
        \STATE $\fvals, \vweights \gets \complex[0]$
        \STATE $\diffwecfs \gets \zeros(m, \numvals)$

        \vspace{1ex}
        \STATE $\vindices \gets \alpha(\fvals)$
        \STATE $\scadd(\vindices^T, \texttt{VertexWeights})
        (\diffwecfs)$

        \vspace{1ex}
        \FOR{$i=1$ \TO $\dim(K)$}
            \STATE \texttt{SimplexVertices, SimplexWeights} $\gets \complex[i]$
            \STATE \texttt{SimpIndices} $\gets \vindices
            [\texttt{SimplexVertices}]$
            \STATE $\msi \gets \rmax(\texttt{SimpIndices},1)$

            \vspace{1ex}
            \STATE $\scadd(\msi^T,
            (-1)^i * \texttt{SimplexWeights}) (\diffwecfs)$
        \ENDFOR

        \vspace{1ex}
        \STATE $\wecfs \gets \cumsum(\diffwecfs)$\\
        \RETURN $\wecfs$
    \end{algorithmic}
\end{algorithm}

Here, we prove that \algref{WECT} produces the tensor $\wecfs$ with shape
$(m, \numvals)$ and~values
\begin{equation}
    \wecfs[p, q] = \wecf_{f_p} \big( \beta(q) \big).
\end{equation}
The algorithm begins by initializing a zero-tensor $\diffwecfs$ with shape
$(m,\numvals)$.
Next, the vertex filter values are mapped to indices in $[[\numvals]]$ by
applying the map $\alpha$ elementwise to the tensor $\fvals$.
This gives the tensor $\vindices$ with values
\begin{equation} \label{eq:v-inds}
    \vindices[a, p] = \alpha \big( f_p(v_a) \big).
\end{equation}
The operation $\scadd(\vindices^T, \texttt{VertexWeights})$ is then applied to
$\diffwecfs$, resulting in adding to $\diffwecfs$ the tensor $D_0$ with shape
$(m, \numvals)$ and values
\[
    D_0[p,q] = \sum_{\substack{a \in [[k_0]] \\ \vindices[a, p] = q}}
    \texttt{VertexWeights}[a].
\]

Next, inside the for-loop, fix an $i \in \{1, 2, \ldots, \dim(K)\}$.
The tensor \texttt{SimplexVertices} has rows indexed by $K_i$ with the entries
in each row corresponding to the vertex sets of simplices in $K_i$.
Therefore, the tensor~$\msi$ has shape~$(k_i, m)$ and values
\begin{align} \label{eq:msi}
    \nonumber
    \msi[b, p] &= \max_{j \in [[i+1]]}
    \vindices \big[ \texttt{SimplexVertices}[b, j], p \big]\\
    \nonumber
    &= \max_{v \in \sigma_b^i} \alpha \big( f_p (v) \big)\\
    \nonumber
    &= \alpha \big( \max_{v \in \sigma_b^i} f_p (v) \big)
    \hspace{1.5cm} \text{as } \alpha \text{ is order-preserving}\\
    &= \alpha \big( f_p^{\max} (\sigma_b^i) \big).
\end{align}
The operation $\scadd(\msi^T, (-1)^i * \texttt{SimplexWeights})$ updates
$\diffwecfs$ by adding to it the tensor $D_i$ with shape
$(m, \numvals)$ and values
\[
    D_i[p,q] = \sum_{\substack{b \in [[k_i]] \\ \msi[b,p] = q}}
    (-1)^i \texttt{SimplexWeights}[b].
\]

After the conclusion of the for loop, we have
$
    \diffwecfs = \sum_{i=0}^{\dim(K)} D_i.
$
Finally, a cumulative sum is applied to $\diffwecfs$ to obtain the tensor
$\wecfs$.
From the fact that the cumulative sum is a linear operator, we have
\[
    \wecfs = \sum_{i = 0}^{\dim(K)} \cumsum(D_i).
\]
The cumulative sum of $D_0$ has values
\begin{align} \label{eq:vertex-sum}
    \nonumber
    \cumsum(D_0)[p,q]
    &= \sum_{q' \leq q} \sum_{\substack{a \in [[k_0]] \\ \vindices[a, p]= q'}}
    \texttt{VertexWeights}[a]\\
    \nonumber
    &= \sum_{\substack{a \in [[k_0]] \\ \vindices[a, p] \leq q}}
    \texttt{VertexWeights}[a]\\
    \nonumber
    &= \sum_{\substack{a \in [[k_0]] \\ \alpha( f_p( v_a ) ) \leq q}}
    \texttt{VertexWeights}[a]
    &\text{by \eqref{v-inds}}\\
    \nonumber
    &= \sum_{\substack{a \in [[k_0]] \\ f_p(v_a) \leq \beta(q)}}
    \texttt{VertexWeights}[a]
    &\text{by \eqref{galois}}\\
    &= \sum_{\substack{ v \in K_0 \\ f_p(v) \leq \beta(q) }}
    \omega(v).
\end{align}
Likewise, for any $i \in \{1, \ldots, \dim(K)\}$, \eqref{msi} together with the
same argument used to show \eqref{vertex-sum} gives
\[
    \cumsum(D_i)[p,q] = \sum_{\substack{\sigma \in K_i \\ f_p^{\max} (\sigma)
    \leq \beta(q)}} (-1)^i \omega(\sigma).
\]
Therefore, the tensor $\wecfs$ has values
\begin{align*}
    \wecfs[p,q] &= \sum_{\substack{ v \in K_0 \\ f_p(v) \leq \beta(q) }}
    \omega(v) + \sum_{i = 1}^{\dim(K)}
    \sum_{\substack{\sigma \in K_i \\ f_p^{\max} (\sigma)
    \leq \beta(q)}} (-1)^i \omega(\sigma)\\
    &= \sum_{\substack{\sigma \in K \\ f_p^{\max}(\sigma) \leq \beta(q)}}
    (-1)^{\dim(\sigma)} \omega(\sigma)\\
    &= \wecf_{f_p} \big( \beta(q) \big).
\end{align*}

\begin{example} \label{ex:wect}
    Computing a discretization of the WECT of a weighted
    simplicial complex~${\K= (K, \omega)}$ with a map $\ell\colon K_0 \to \R^n$
    is a special case of \algref{WECT} as follows.
    Consider a tensor $D$, with shape $(d, n)$, of direction vectors over which
    to sample the WECT.
    For any~$p \in [[d]]$, let $\vect s_{p} \in \R^n$ be the $p$-th row of $D$.
    Let~$V$ be the tensor, with shape $(k_0, n)$, of the vertex coordinates of
    $K$.
    Then, the tensor $V * D^T$ has values
    \[
        (V * D^T)[a, p] = h_{\vect s_p}(v_a).
    \]
    Therefore, using the tensor $V * D^T$ as the $\fvals$ tensor in
    \algref{WECT} produces a discretization of the WECT over the finite sample
    of directions in $D$.
\end{example}

\subsection{Complexity Analysis}

Here, we analyze the asymptotic complexity of \algref{WECT}.
Assume that the ambient dimension $n$ is fixed.
Let $k$ be the total number of simplices of $K$, let $m$ be the number of vertex
filters, and let $\numvals$ be the number of values to sample the vertex
filters~over.

The input $\complex$ requires $\Theta(k m)$ memory usage.
The algorithm first initializes a zero tensor $\diffwecfs$ of
shape~$(m,\numvals)$, requiring~$\Theta(m \numvals)$ memory.
The function $\alpha$ (see \eqref{left-adjoint}) is then applied elementwise to
$\fvals$, taking~$\Theta(k m)$ time.

Next, the operation
$
\scadd(\vindices^T, \texttt{VertexWeights})
$
is applied to $\diffwecfs$.
ScatterAdd adds a value from \texttt{VertexWeights} to $\diffwecfs$ for each
entry of $\vindices^T$, resulting in a time complexity
of~$\Theta(k m)$ and no additional memory usage as the operation
is performed in place.

Now, for a fixed dimension $i \in \{1, 2, \ldots n\}$, the tensor
\[
    \texttt{SimpIndices} = \vindices[\texttt{SimplexVertices}]
\]
with shape $(k_i, i+1, m)$ is constructed,
taking $\Theta(k m)$ time and memory.
This tensor is then reduced along the second axis by taking the maximum values,
creating the tensor $\msi$ with shape $(k_i, m)$ in~$\Theta(k m)$ time and memory.
The tensor $\diffwecfs$ is then updated by applying the map
\[
    \scadd(\msi^T, (-1)^i * \texttt{SimplexWeights})
\]
which takes $\Theta(k m)$ time and memory.

Thus, we have a total time and memory complexity of
$
    \Theta ( n k m )
    = \Theta ( k m )
$
after conclusion of the for-loop.
Finally, a cumulative sum is applied to $\diffwecfs$, taking
$\Theta(m \numvals)$ time and using no additional memory.
This results in
$\Theta(m (k + \numvals))$ for \algref{WECT}
as the final time and memory complexity.
In practice, $m$ and $\numvals$ are typically held constant.
Indeed, this is necessary to achieve consistent vectorizations of data.
Holding $m$ and $\numvals$ constant results in a
final time and memory complexity of~$\Theta(k)$ for \algref{WECT}.

\section{Experiments}

In order to demonstrate the efficacy of our implementation,
we evaluate and compare the runtime of Algorithm~\ref{alg:WECT} with existing
methods (see \secref{existing-frameworks}) on a variety of~datasets.\footnote{
    The implementation of \pyect can be accessed here:
    \url{https://anonymous.4open.science/r/pyECT-anonymous-902F/}.
    The source code for reproducing experiments (with instructions)
    is available here:
    \url{https://anonymous.4open.science/r/pyECT-experiments-anonymous-0E4F/}.
}

\subsection{Experimental Design}
Experiments were performed on a single node with a single CPU or a single GPU
(unless otherwise specified).
EPYC Rome CPUs and NVIDIA A100 GPUs were used.
Runtimes for CPU experiments were recorded using the \texttt{time} Python
library using the performance counter for benchmarking.
For the GPU experiments, runtimes were recorded using \texttt{PyTorch}'s
\texttt{torch.cuda.Event} API to prevent inaccurate timing from
asynchronous GPU operations.
The methods were evaluated on the following two- and three-dimensional datasets:
\begin{itemize}
    \item \textbf{Fashion-MNIST} \cite{xiaoFashionMNIST}.
        Fashion-MNIST is a dataset consisting of $28 \times 28$ grayscale
        images of clothing items.
        Its training partition consists of $60000$ images with equal
        representation across ten classes ($6000$ images per class).
        In our experiments, we use a random subset of $1000$ images sampled
        uniformly without replacement from the training partition, which are
        then converted to cubical complexes.

    \item \textbf{ImageNet} \cite{deng2009ImageNet, schwartz2020ImagenetSample}.
        The ImageNet Sample Image dataset~\cite{schwartz2020ImagenetSample},
        which consists of $1000$ samples from the Large Scale Visual
        Recognition Challenge (ILSVRC) 2012-2017 subset of the
        ImageNet dataset~\cite{deng2009ImageNet} was used.
        All images in the sample were converted to grayscale using the
        ITU-R 601-2 luma transform within the Pillow Python
        library~\cite{Pillow2024}.
        Then, the images were padded to the maximum dimensions of all images
        in the dataset, resulting in each image having uniform size of
        $1632 \times 2400$ pixels.
        The padded pixels were assigned a pixel intensity value of $1.0$
        (on a zero to one scale), and the images were converted to cubical
        complexes.

    \item \textbf{3DCCs}. A synthetic dataset of $1000$ three-dimensional
        cubical complexes was generated. Each complex has dimension
        $100 \times 100 \times 100$, with pixel intensities drawn
        i.i.d.\ from the uniform distribution on $[0,1]$.

    \item \textbf{Stanford Bunny and Armadillo} \cite{stanford3dscanrepo}.
        The Stanford 3D Scanning Repository provides
        triangulated mesh models of three-dimensional objects.
        We use the Bunny and Armadillo models.
        The meshes were converted to (unweighted) simplicial complexes after
        preprocessing to de-duplicate vertices, edges, and faces.
        In the resulting simplicial complexes, the Bunny dataset consists of
        $2503$ vertices, $7473$ edges, and $4968$ faces.
        The Armadillo is significantly larger, consisting of $172974$ vertices,
        $518916$ edges, and $345944$ faces.

\end{itemize}

\subsection{Results}
We now present the experimental results, measured by the average speedup factor
of our method over existing methods.
The average speedup factor is defined as the runtime of the baseline method
(e.g., \eucalc, \dect, or \fasttopology)
divided by the runtime of the proposed method, \pyect,
averaged across all images in the dataset.
Across all figures, error bars represent $95\%$ confidence intervals.

\paragraph*{WECT Results}
In this section, we present results on the average speedup of \pyect
over \eucalc and \dect when computing the WECT of the Fashion-MNIST and
ImageNet datasets.
As \eucalc does not provide a GPU implementation, all results for \eucalc
are measured on the CPU.
Also, as \eucalc does not natively provide a vectorized result,
we provide results both including and excluding the vectorization time.

\begin{figure}[htbp]
    \centering
    \begin{subfigure}[t]{.48\textwidth}
        \centering
        \includegraphics[width=\textwidth]{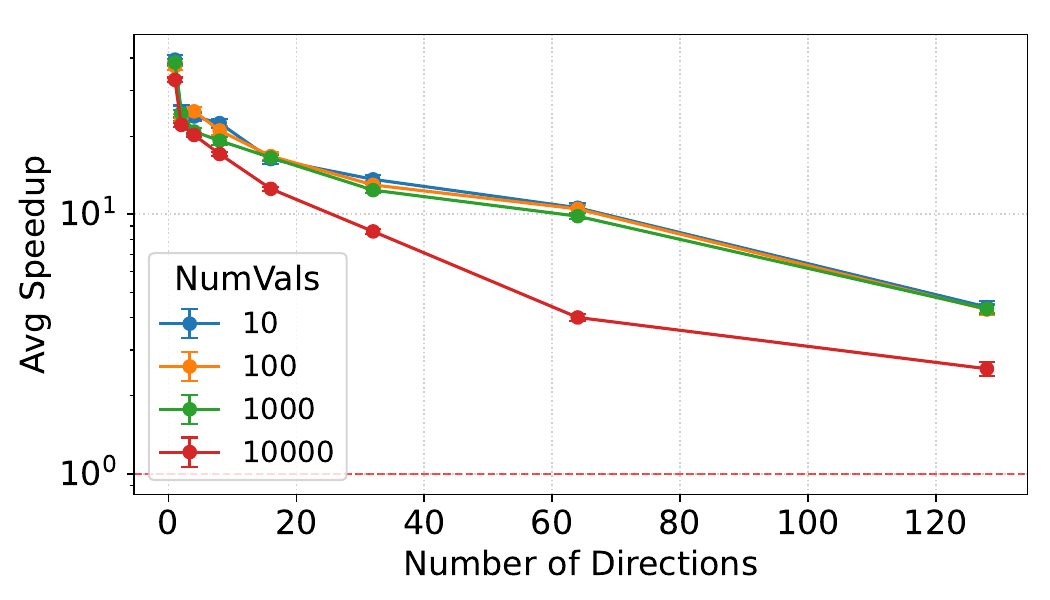}
        \subcaption{\pyect (CPU) vs.\ \eucalc}
        \label{fig:exp_wect_fmnist-eucalc_v_pyect_cpu}
    \end{subfigure}%
    ~
    \begin{subfigure}[t]{.48\textwidth}
        \centering
        \includegraphics[width=\textwidth]{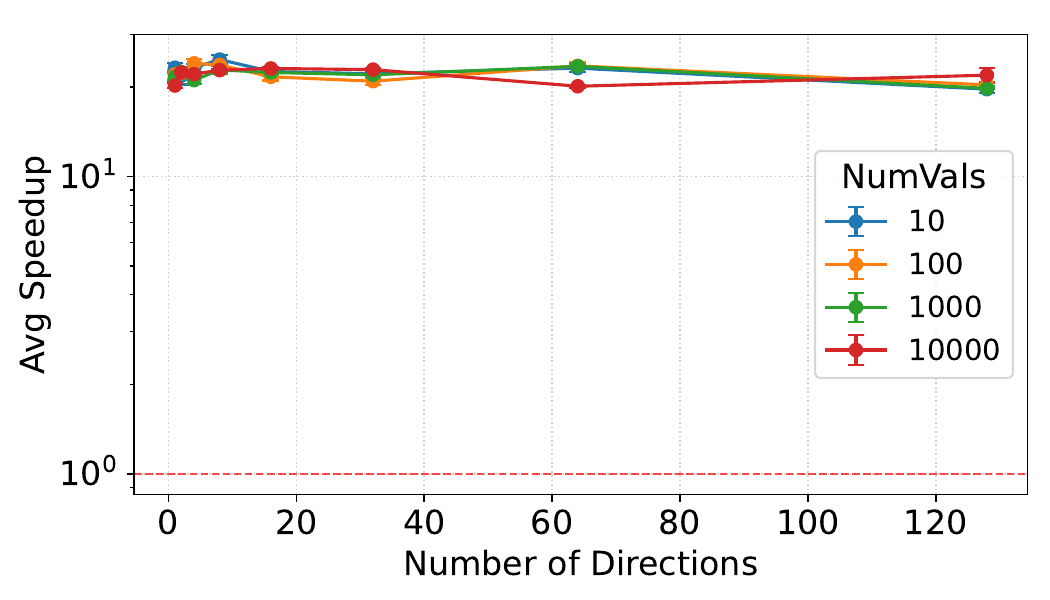}
        \subcaption{\pyect (GPU) vs.\ \eucalc}
        \label{fig:exp_wect_fmnist-eucalc_v_pyect_cuda}
    \end{subfigure}%
    \\
    \begin{subfigure}[t]{.48\textwidth}
        \centering
        \includegraphics[width=\textwidth]{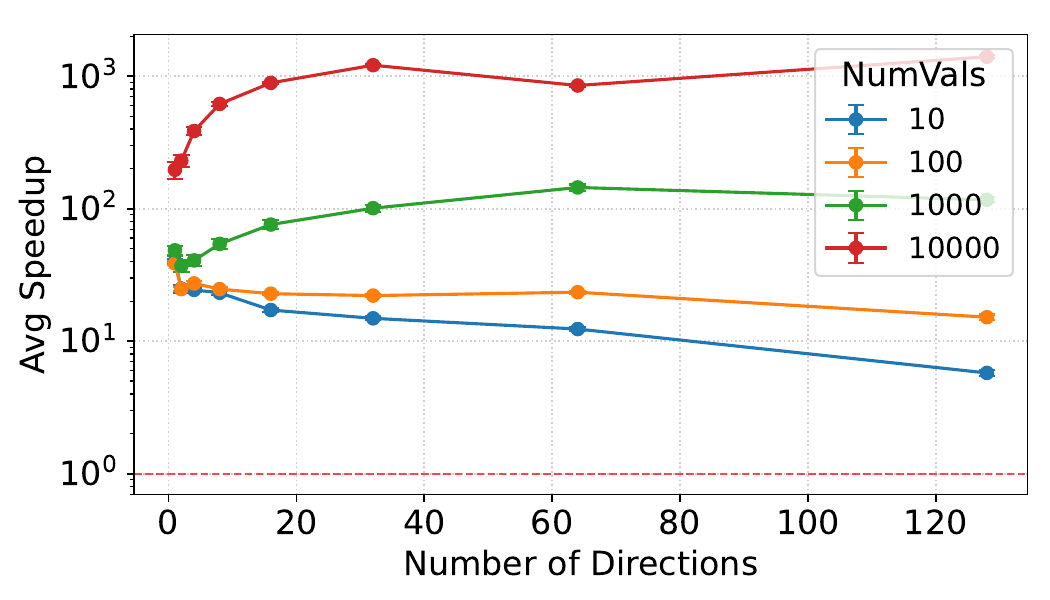}
        \subcaption{\pyect (CPU) vs.\ \eucalc (vectorized)}
        \label{fig:exp_wect_fmnist-eucalc_vect_v_pyect_cpu}
    \end{subfigure}
    ~
    \begin{subfigure}[t]{.48\textwidth}
        \centering
        \includegraphics[width=\textwidth]{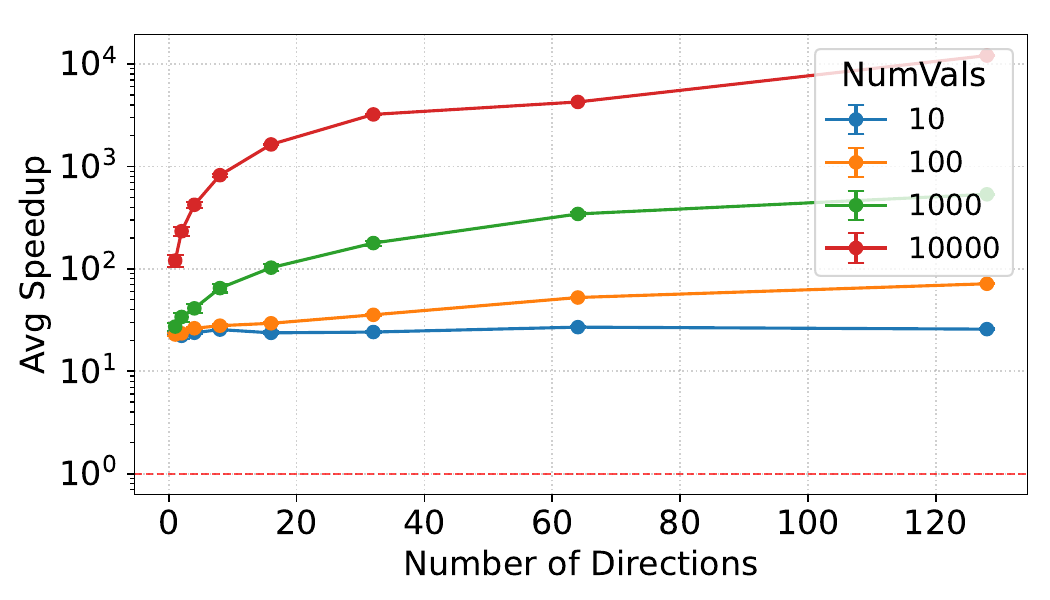}
        \subcaption{\pyect (GPU) vs.\ \eucalc (vectorized)}
        \label{fig:exp_wect_fmnist-eucalc_vect_v_pyect_cuda}
    \end{subfigure}
    \\
    \begin{subfigure}[t]{.48\textwidth}
        \centering
        \includegraphics[width=\textwidth]{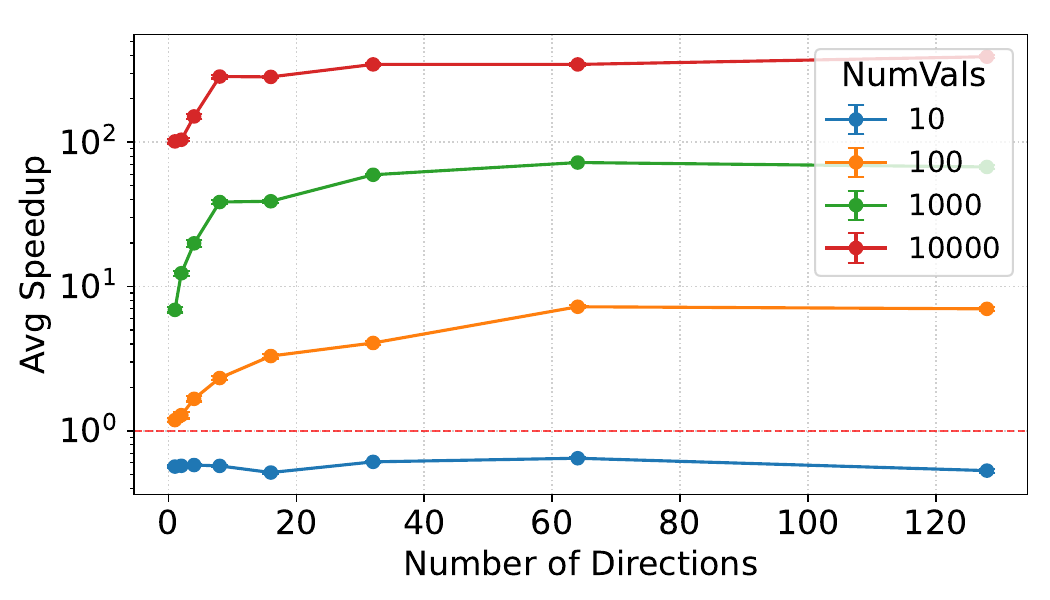}
        \subcaption{\pyect (CPU) vs.\ \dect (CPU)}
        \label{fig:exp_wect_fmnist-dect_v_pyect_cpu}
    \end{subfigure}
    ~
    \begin{subfigure}[t]{.48\textwidth}
        \centering
        \includegraphics[width=\textwidth]{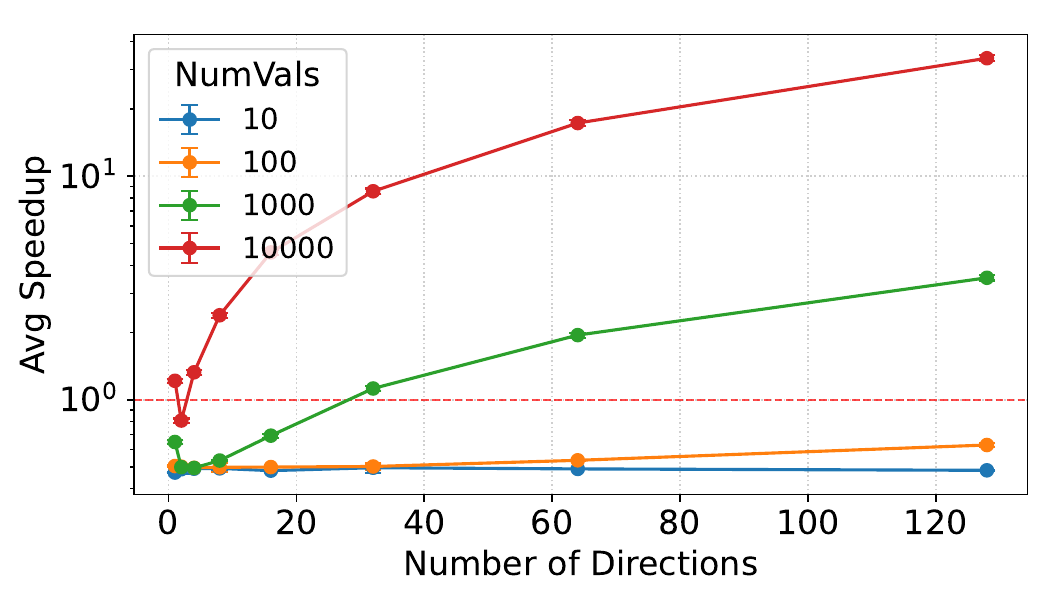}
        \subcaption{\pyect (GPU) vs.\ \dect (GPU)}
        \label{fig:exp_wect_fmnist-dect_v_pyect_cuda}
    \end{subfigure}
    \caption{
        A comparison for computing
        on the Fashion-MNIST dataset for WECT computation. Figures
        \subrefletter{fig:exp_wect_fmnist-eucalc_v_pyect_cpu} and
        \subrefletter{fig:exp_wect_fmnist-eucalc_v_pyect_cuda} compare
        \pyect with \eucalc excluding the vectorization time,
        \subrefletter{fig:exp_wect_fmnist-eucalc_vect_v_pyect_cpu} and
        \subrefletter{fig:exp_wect_fmnist-eucalc_vect_v_pyect_cuda} compare
        \pyect with \eucalc including the vectorization time,
        and \subrefletter{fig:exp_wect_fmnist-dect_v_pyect_cpu} and
        \subrefletter{fig:exp_wect_fmnist-dect_v_pyect_cuda} compare \pyect with \dect.
        CPU implementations of \pyect are shown on the left
        and GPU implementations are shown on the right.
        Vertical axes are not consistent across subfigures.
        In nearly all cases, \pyect outperforms the baseline methods.
        This behavior is especially pronounced when using the GPU architecture
        with large numbers of directions and filtration heights.
        } \label{fig:exp_wect_fmnist}
\end{figure}

The results for Fashion-MNIST are shown in \figref{exp_wect_fmnist}.
Across all settings, \pyect demonstrates speedups over both
\eucalc and \dect.
When comparing against \eucalc
(\figref{exp_wect_fmnist}(\subref{fig:exp_wect_fmnist-eucalc_v_pyect_cpu})--(\subref{fig:exp_wect_fmnist-eucalc_vect_v_pyect_cuda})),
\pyect demonstrates a speedup in both CPU and GPU
implementations---even without the vectorization
step.
However, the GPU implementation of \pyect significantly outperforms the CPU
implementation as the number of directions increases, which is especially
pronounced when including the vectorization time for \eucalc.
As expected, when increasing the number of values used in the filtration,
the speedup factor of \pyect over \eucalc increases drastically.
When comparing against \dect
(subfigures (\subref{fig:exp_wect_fmnist-dect_v_pyect_cpu}) and
(\subref{fig:exp_wect_fmnist-dect_v_pyect_cuda})),
\pyect also demonstrates significant speedups.
While the computation speed is comparable for low numbers of directions and
filtration heights, \pyect significantly outperforms \dect for larger choices
of parameters.
Also, as \dect is optimized for GPU architectures, the average speedup on
GPU architectures is less pronounced than that of other comparisons.
However, when constrained to CPU architectures, \pyect is much faster,
indicating another advantage of \pyect's general applicability for all
scenarios.

\begin{figure}[htbp]
    \centering
    \begin{subfigure}[t]{.48\textwidth}
        \centering
        \includegraphics[width=\textwidth]{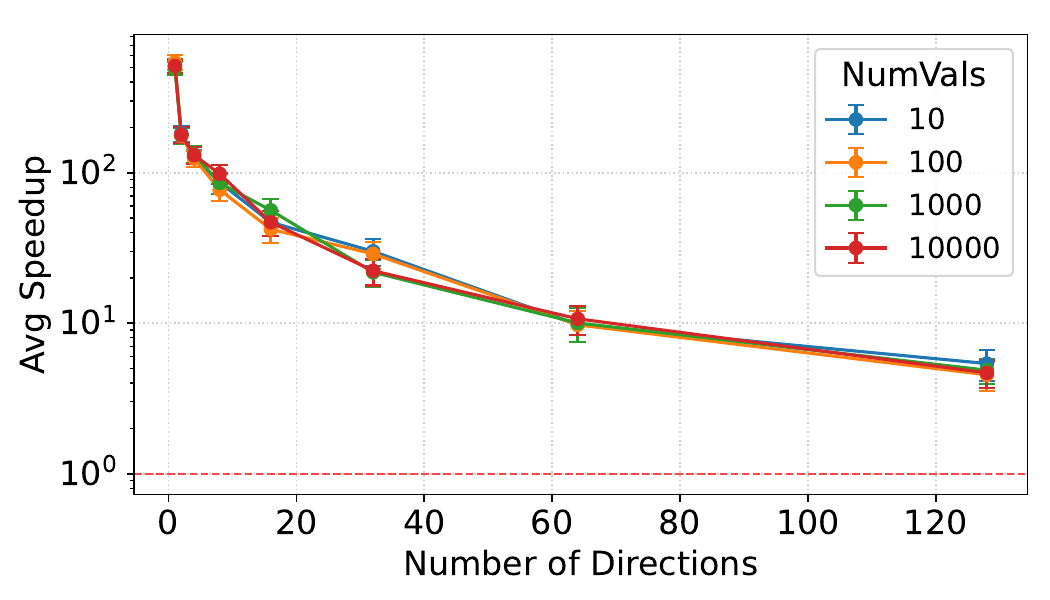}
        \subcaption{\pyect (CPU) vs.\ \eucalc}
        \label{fig:exp_wect_imagenet-eucalc_v_pyect_cpu}
    \end{subfigure}%
    ~
    \begin{subfigure}[t]{.48\textwidth}
        \centering
        \includegraphics[width=\textwidth]{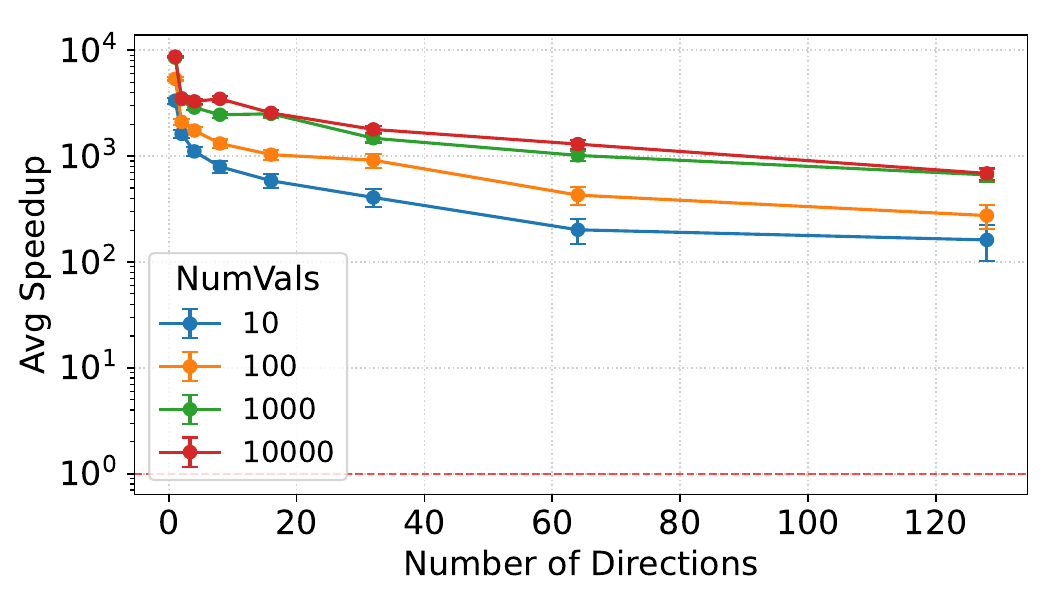}
        \subcaption{\pyect (GPU) vs.\ \eucalc}
        \label{fig:exp_wect_imagenet-eucalc_v_pyect_cuda}
    \end{subfigure}%
    \\
    \begin{subfigure}[t]{.48\textwidth}
        \centering
        \includegraphics[width=\textwidth]{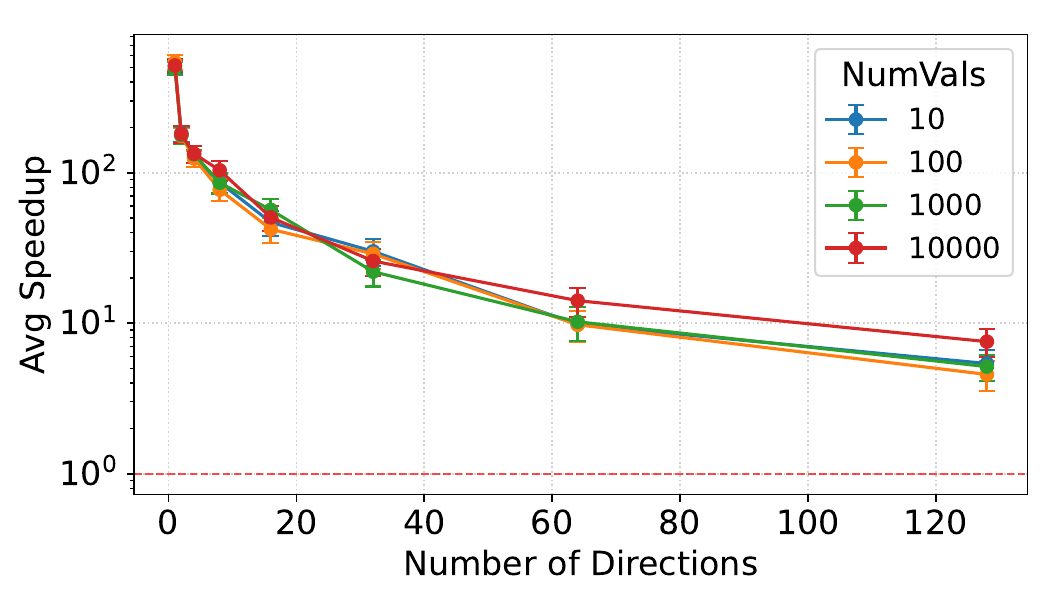}
        \subcaption{\pyect (CPU) vs.\ \eucalc (vectorized)}
        \label{fig:exp_wect_imagenet-eucalc_vect_v_pyect_cpu}
    \end{subfigure}
    ~
    \begin{subfigure}[t]{.48\textwidth}
        \centering
        \includegraphics[width=\textwidth]{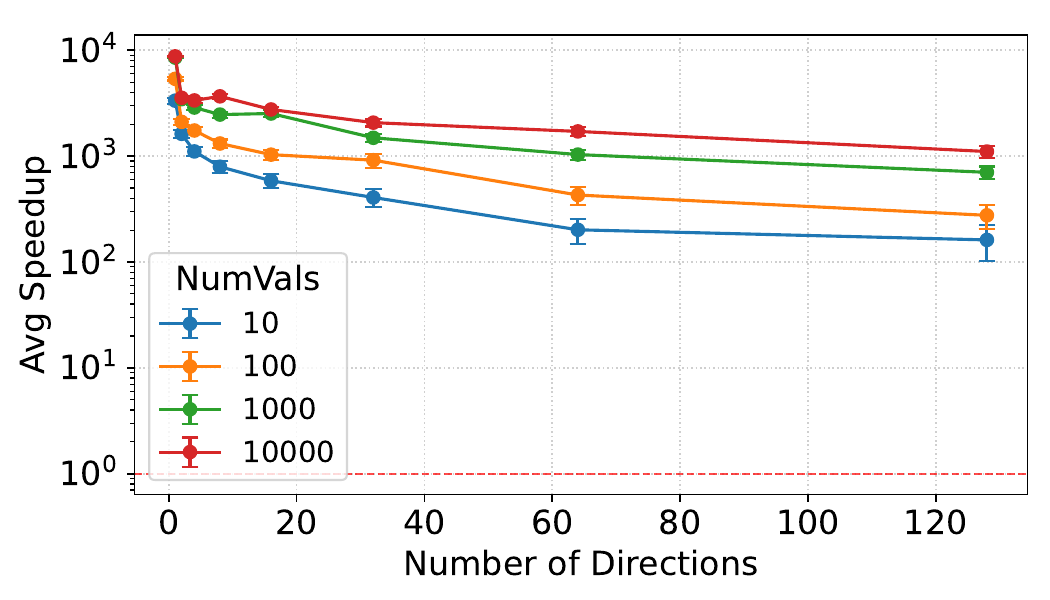}
        \subcaption{\pyect (GPU) vs.\ \eucalc (vectorized)}
        \label{fig:exp_wect_imagenet-eucalc_vect_v_pyect_cuda}
    \end{subfigure}
    \\
    \begin{subfigure}[t]{.48\textwidth}
        \centering
        \includegraphics[width=\textwidth]{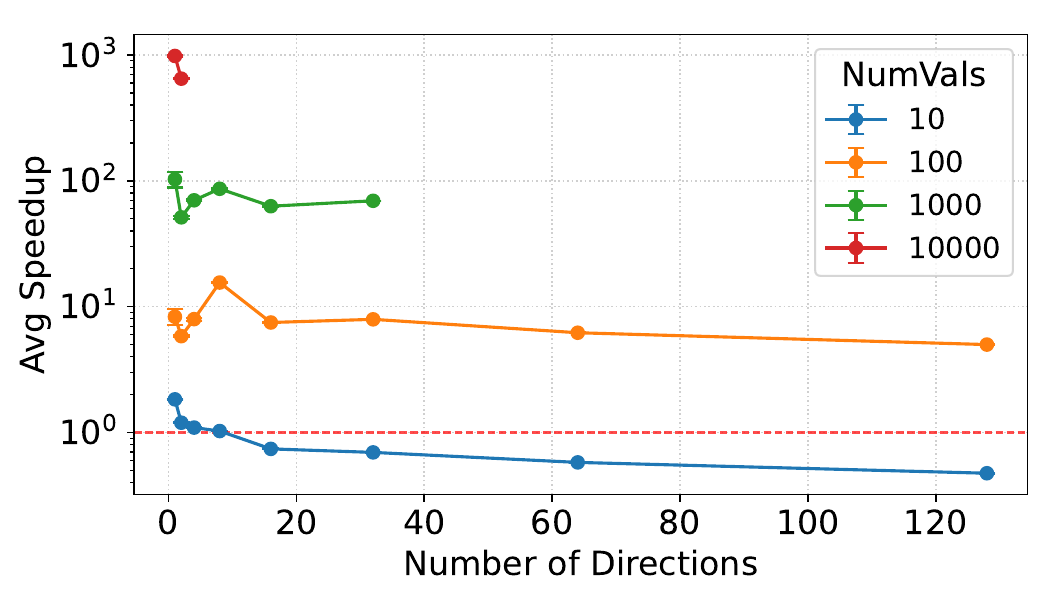}
        \subcaption{\pyect (CPU) vs.\ \dect (CPU)}
        \label{fig:exp_wect_imagenet-dect_v_pyect_cpu}
    \end{subfigure}
    ~
    \begin{subfigure}[t]{.48\textwidth}
        \centering
        \includegraphics[width=\textwidth]{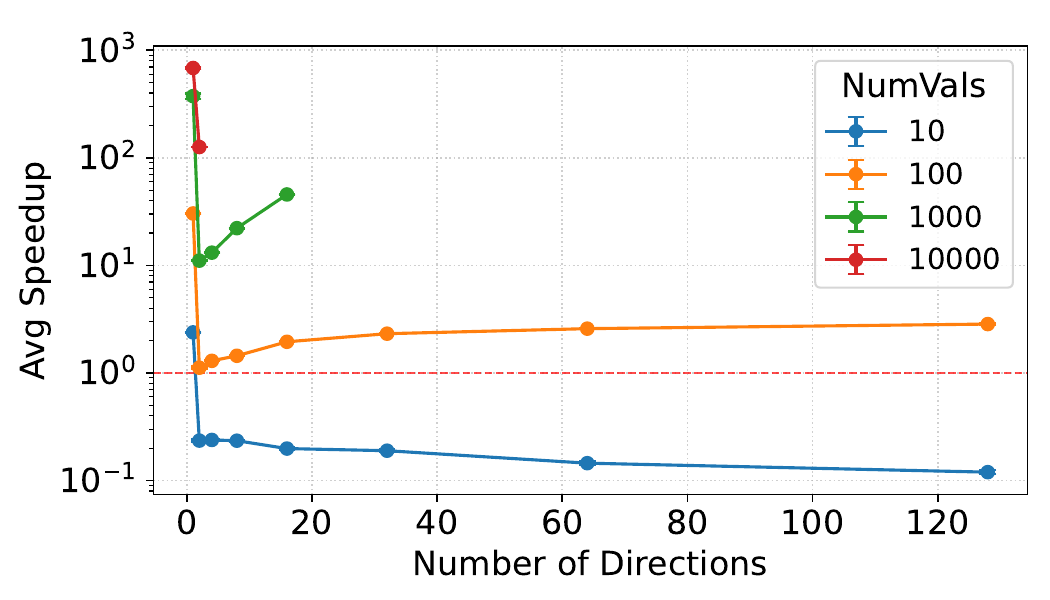}
        \subcaption{\pyect (GPU) vs.\ \dect (GPU)}
        \label{fig:exp_wect_imagenet-dect_v_pyect_cuda}
    \end{subfigure}
    \caption{
        A comparison for computing
        on the ImageNet dataset for WECT computation. Figures
        \subrefletter{fig:exp_wect_imagenet-eucalc_v_pyect_cpu} and
        \subrefletter{fig:exp_wect_imagenet-eucalc_v_pyect_cuda} compare
        \pyect with \eucalc excluding the vectorization time,
        \subrefletter{fig:exp_wect_imagenet-eucalc_vect_v_pyect_cpu} and
        \subrefletter{fig:exp_wect_imagenet-eucalc_vect_v_pyect_cuda} compare
        \pyect with \eucalc including the vectorization time,
        and \subrefletter{fig:exp_wect_imagenet-dect_v_pyect_cpu} and
        \subrefletter{fig:exp_wect_imagenet-dect_v_pyect_cuda} compare
        \pyect with \dect.
        Note that vertical axes are not consistent across subfigures.
        \dect was unable to compute the WECT for large numbers of
        directions and heights due to memory constraints, resulting in missing
        values in \subrefletter{fig:exp_wect_imagenet-dect_v_pyect_cpu} and
        \subrefletter{fig:exp_wect_imagenet-dect_v_pyect_cuda}.
        However, \pyect was able to compute the WECT in these scenarios.
        } \label{fig:exp_wect_imagenet}
\end{figure}

The average speedup results on the ImageNet are shown in
\figref{exp_wect_imagenet}.
When comparing against \eucalc (both with and without the vectorization step,
in Figures~\ref{fig:exp_wect_imagenet-eucalc_v_pyect_cpu}--\ref{fig:exp_wect_imagenet-eucalc_vect_v_pyect_cuda}),
\pyect demonstrates a significant speedup in both CPU and GPU implementations.
This speedup is especially pronounced for the GPU implementation of \pyect,
as it is much more stable to large numbers of directions and filtration
height values.
Due to the large size of the complexes in this dataset, the CPU speedup of
\pyect is less pronounced than on the Fashion-MNIST dataset (especially
for large numbers of directions and heights).
Furthermore, the vectorization time of \eucalc is less impactful on this dataset,
as the vectorization step is less significant compared to the overall runtime
for the large complexes.
In the GPU implementation of \pyect, we observe a trend where the average
speedup increases as the number of height filtration values increases.
While this behavior appears counterintuitive, in reality the vectorized code
encounters less efficient memory access patterns at lower numbers of heights,
which leads to a smaller advantage over \eucalc with the smaller tensor values.
For a technical description of memory access patterns specific to
the GPU hardware used in these experiments, we refer the reader to
\cite{gpu2025}.
In the comparison against \dect (Figures~\ref{fig:exp_wect_imagenet-dect_v_pyect_cpu} and
\ref{fig:exp_wect_imagenet-dect_v_pyect_cuda}), \pyect again demonstrates significant
speedups.
In fact, due to the large size of the complexes in this dataset,
\dect was unable to compute the WECT for large numbers of directions
and height values (due to memory constraints on the GPU).
This results in the appearance of missing values in
Figures \ref{fig:exp_wect_imagenet-dect_v_pyect_cpu} and
\ref{fig:exp_wect_imagenet-dect_v_pyect_cuda}.
Note, however, that \pyect is still able to compute the WECT in these
scenarios, demonstrating its memory efficiency.
Furthermore, in the cases where the WECT could be computed by \dect,
the \pyect implementation outperforms \dect for larger numbers of directions
and height filtration values.

\begin{figure}[htbp]
    \centering
    \begin{subfigure}[t]{.48\textwidth}
        \centering
        \includegraphics[width=\textwidth]{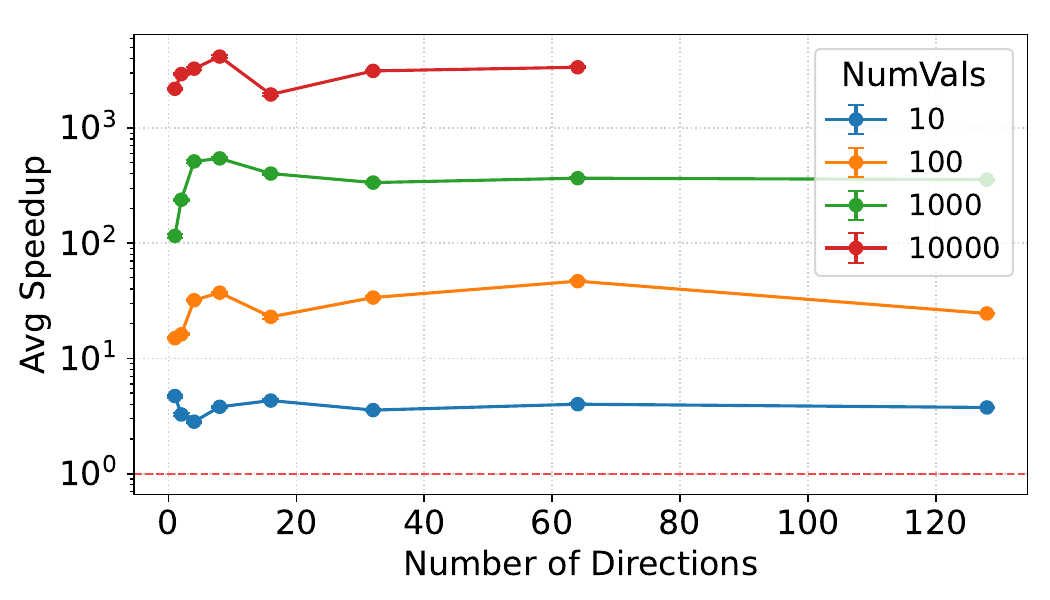}
        \subcaption{Bunny: \pyect (CPU) vs.\ \dect (CPU)}
        \label{fig:exp_wect_bunny-dect_v_pyect_cpu}
    \end{subfigure}%
    ~
    \begin{subfigure}[t]{.48\textwidth}
        \centering
        \includegraphics[width=\textwidth]{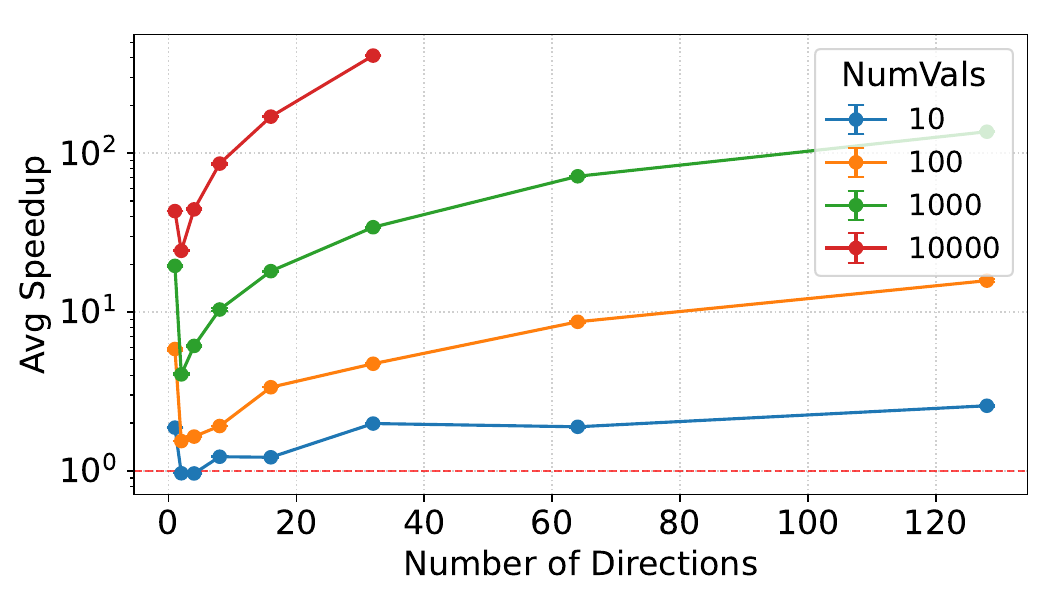}
        \subcaption{Bunny: \pyect (GPU) vs.\ \dect (GPU)}
        \label{fig:exp_wect_bunny-dect_v_pyect_cuda}
    \end{subfigure}%
    \\
    \begin{subfigure}[t]{.48\textwidth}
        \centering
        \includegraphics[width=\textwidth]{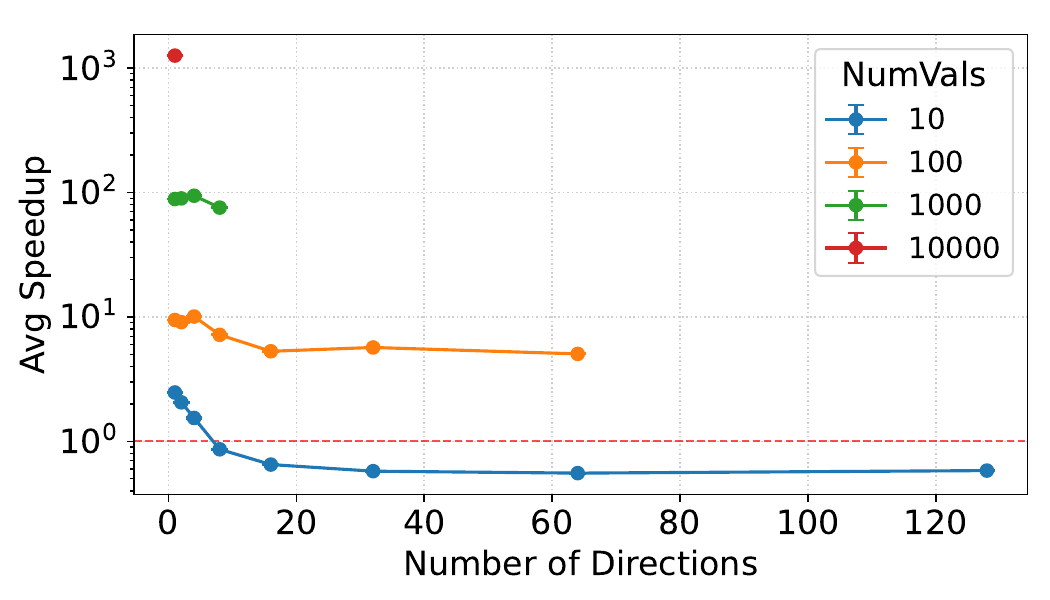}
        \subcaption{Armadillo: \pyect (CPU) vs.\ \dect (CPU)}
        \label{fig:exp_wect_armadillo-dect_v_pyect_cpu}
    \end{subfigure}
    ~
    \begin{subfigure}[t]{.48\textwidth}
        \centering
        \includegraphics[width=\textwidth]{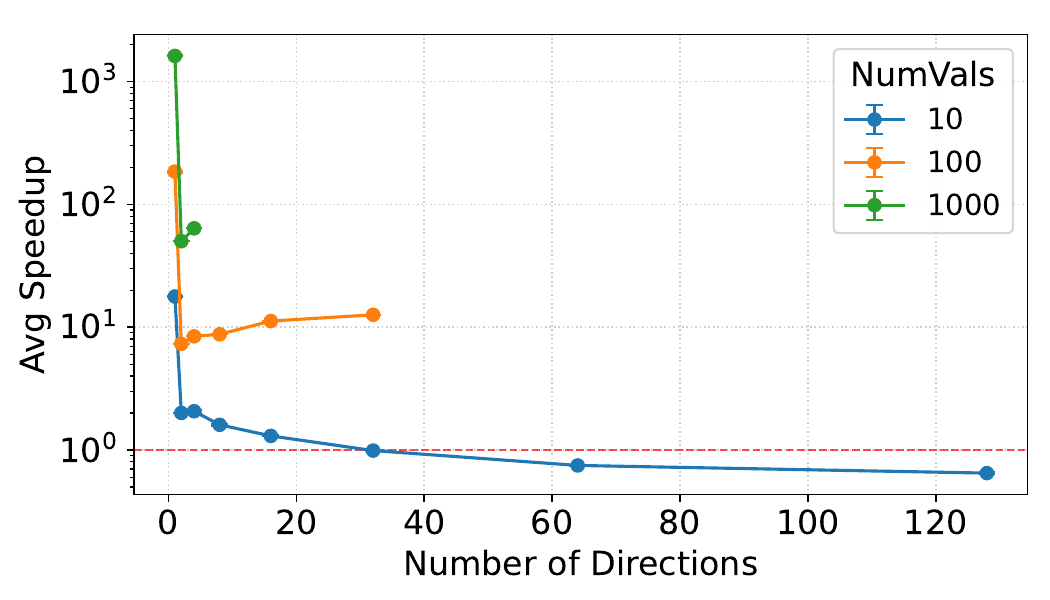}
        \subcaption{Armadillo: \pyect (GPU) vs.\ \dect (GPU)}
        \label{fig:exp_wect_armadillo-dect_v_pyect_cuda}
    \end{subfigure}
    \caption{
        A comparison of \pyect against \dect for computing the ECT on
        the Stanford Bunny and Armadillo meshes.
        Note that vertical axes are not consistent across subfigures.
        In many cases, \dect was unable to compute the ECT due to
        memory constraints (leading to missing values in the plots).
        However, \pyect was able to compute the ECT in all scenarios,
        and especially outperforms \dect for large numbers of directions and
        filtration heights.
        } \label{fig:exp_wect_mesh}
\end{figure}

Speedup results for the comparison of \pyect against \dect
on the three-dimensional mesh datasets are presented in \figref{exp_wect_mesh}.
Note that experiments were run 10 times on each mesh dataset to generate
the average and confidence interval values shown in the plots.
For large choices of directions and filtration heights in both datasets,
\dect was unable to compute the ECT due to memory constraints.
While this results in missing values in all subplots, from the cases
where \dect was able to compute the ECT,
\pyect outperforms \dect (especially for larger choices of parameters).
The memory issues are more pronounced in the Armadillo dataset
(due to its larger size).
Note that while \pyect demonstrates significant speedups in the GPU
implementations, its advantage is even more pronounced in the CPU
implementations, highlighting the versatility of \pyect across
different hardware architectures.

\begin{figure}[htbp]
    \centering
    \begin{subfigure}[t]{.48\textwidth}
        \centering
        \includegraphics[width=\textwidth]{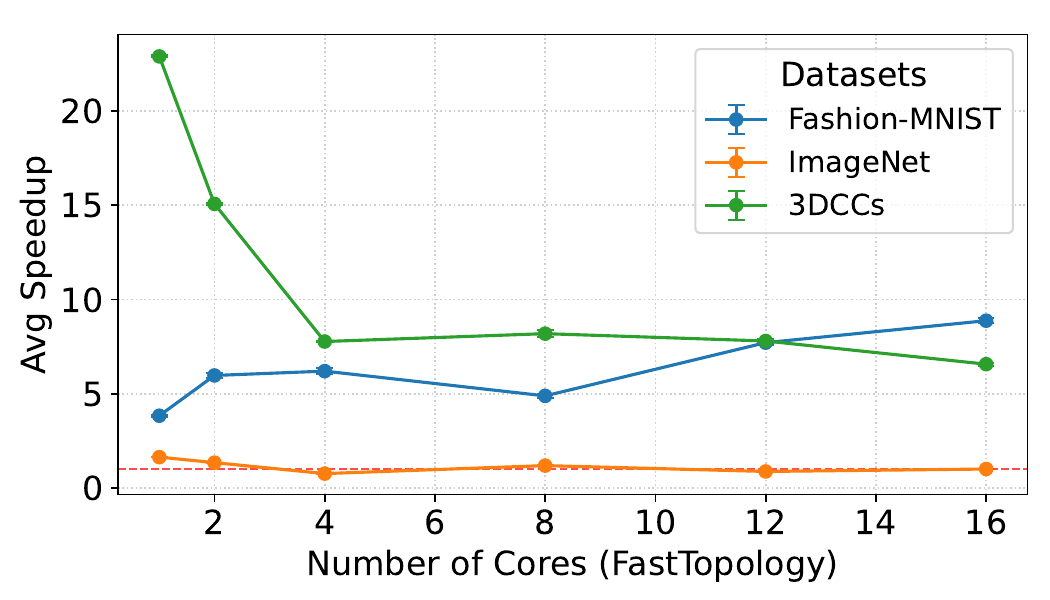}
        \subcaption{\pyect (CPU) vs.\ \fasttopology}
        \label{fig:exp_ecf-ft_v_pyect_cpu}
    \end{subfigure}%
    ~
    \begin{subfigure}[t]{.48\textwidth}
        \centering
        \includegraphics[width=\textwidth]{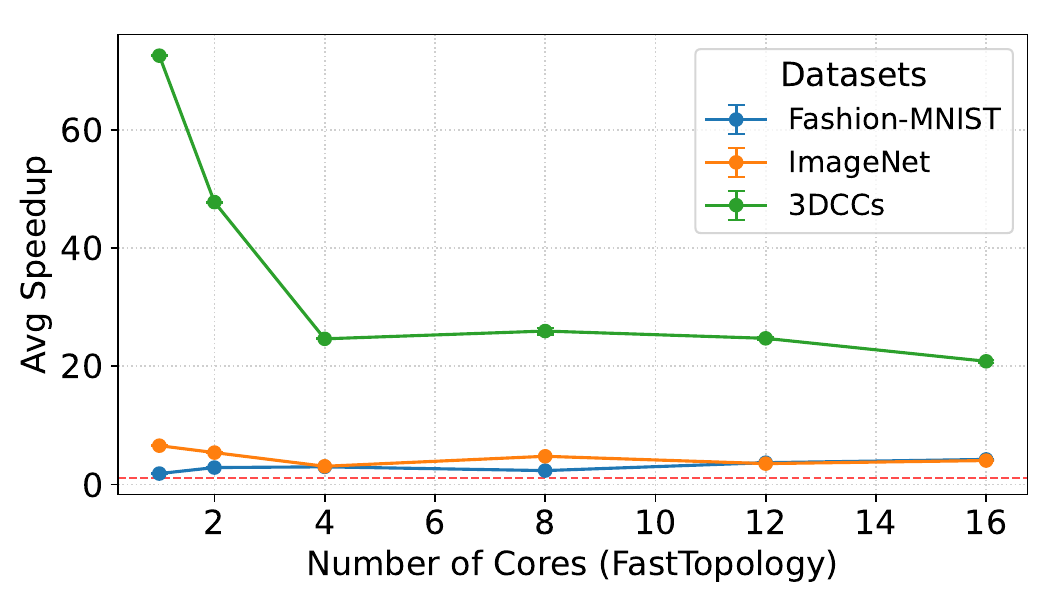}
        \subcaption{\pyect (GPU) vs.\ \fasttopology}
        \label{fig:exp_ecf-ft_v_pyect_cuda}
    \end{subfigure}%
    \\
    \begin{subfigure}[t]{.48\textwidth}
        \centering
        \includegraphics[width=\textwidth]{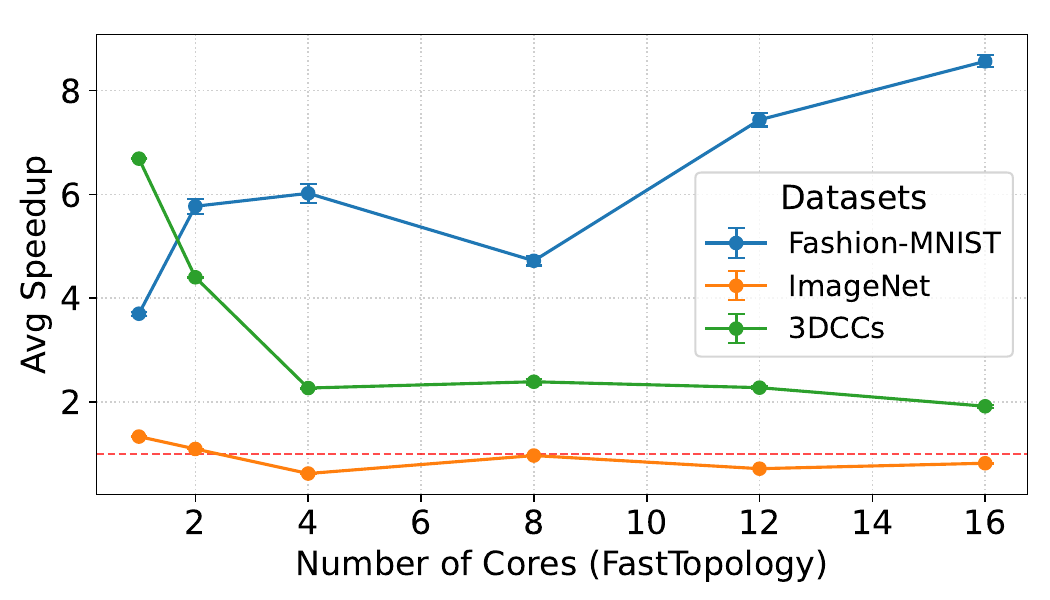}
        \subcaption{\pyect (compiled, CPU) vs.\ \fasttopology}
        \label{fig:exp_ecf-ft_v_pyect_compiled_cpu}
    \end{subfigure}
    ~
    \begin{subfigure}[t]{.48\textwidth}
        \centering
        \includegraphics[width=\textwidth]{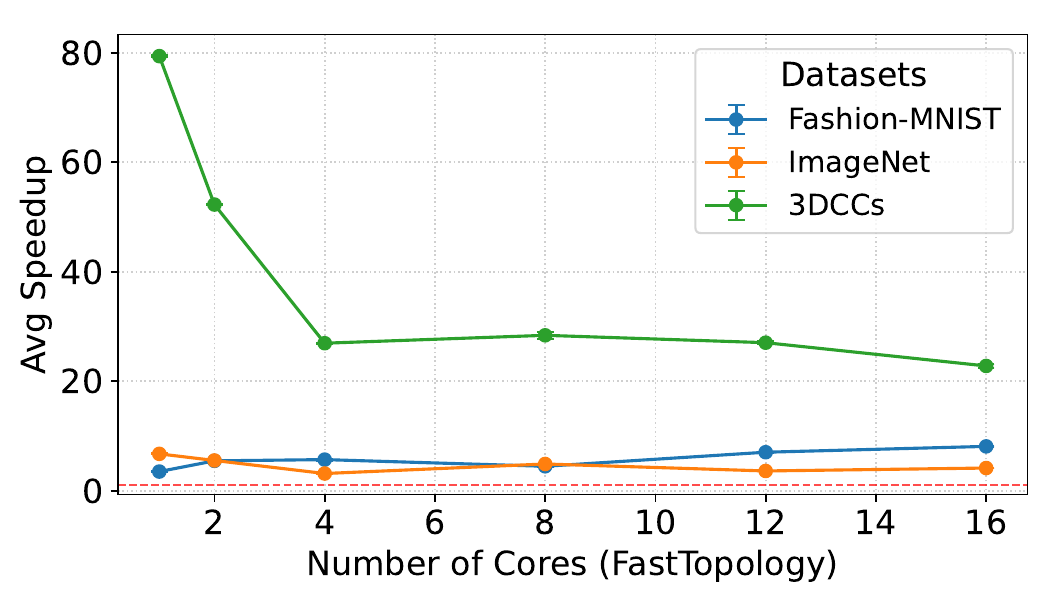}
        \subcaption{\pyect (compiled, GPU) vs.\ \fasttopology}
        \label{fig:exp_ecf-ft_v_pyect_compiled_cuda}
    \end{subfigure}
    \caption{
        A comparison of average speedup of \pyect over \fasttopology when
        computing the ECF of multiple datasets.
        The comparison evaluates both uncompiled and compiled versions
        of \pyect.
        Note that vertical axes are not consistent across subfigures.
        } \label{fig:exp_ecf}
\end{figure}

\paragraph*{ECF Results}
In this section, we present the speedup of \pyect over
\fasttopology when computing the ECF of the Fashion-MNIST, ImageNet, and 3DCCs
datasets.
Due to the consistent computation graph structure of the ECF algorithm,
\pyect is able leverage the PyTorch TorchDynamo compilation framework.
These compiled runtimes are also measured and compared with \fasttopology.
As \fasttopology is parallelizable across multiple CPU cores,
we include results for \fasttopology using $1$, $2$, $4$, $8$, $12$, and $16$
CPU cores.
Note that \fasttopology does not provide a GPU implementation.
The results are presented in \figref{exp_ecf}.
We observe that \pyect provides a significant speedup over \fasttopology
in nearly all cases.
When constrained to CPU architectures, \pyect has a less pronounced speedup,
but is still comparable in speed on a single CPU core when compared with
the fully parallelized \fasttopology implementation.
When utilizing GPU architectures, \pyect demonstrates a significant speedup
over the fully parallelized \fasttopology, especially on the three-dimensional
3DCCs dataset.
The compilation strategy further increases the speed of \pyect
on GPU architectures, but adds unrewarded overhead on CPU architectures.

Next, to verify the efficiency of the  computation on the GPU,
we compute the ECF of 2D and 3D images with intensity drawn uniformly
and measure throughput as in \cite{wang2023gpu,
wang2023gpu_conference}.
For 2D images (sizes $2048^2$, $4096^2$, and $8192^2$) with data
pre-loaded into GPU memory, uncompiled \pyect achieves
$1.11$--$1.13$~GPix/s at $1{,}000$ thresholds and
$2.40$--$2.52$~GPix/s at $10{,}000$ thresholds,
while the compiled version achieves
$1.28$--$1.32$~GPix/s and~$2.52$--$4.11$~GPix/s, respectively.
For 3D images (sizes $128^3$, $256^3$, and $512^3$), uncompiled \pyect achieves
$0.48$--$0.49$~GVox/s at $1{,}000$ thresholds and
$1.05$--$1.09$~GVox/s at $10{,}000$ thresholds,
while the compiled version achieves
$0.43$--$0.56$~GVox/s and $1.09$--$1.76$~GVox/s,~respectively.

\section{Discussion}

In this work, we present a tensor-based framework for computing WECFs
(including ECFs and the WECT).
This framework allows for efficient computation of these functions using tensor
operations, which are well-suited for modern GPU hardware.
Furthermore, our method works in full generality for weighted simplicial
complexes (or cubical complexes) of arbitrary dimension.
Our framework is implemented in a publicly available Python package
called \pyect.
Experimentally, \pyect demonstrates a significant speedup over
existing methods across a variety of datasets.

The \pyect package also contains a variety of tools for constructing and
working with simplicial/cubical complexes.
Our package includes fully vectorized methods for building weighted Freudenthal
complexes and weighted cubical complexes from images.
Additionally, \pyect has dedicated methods for computing
ECFs of images and WECTs of weighted geometric simplicial complexes in $\R^n$.
These two special cases are implemented as \texttt{PyTorch} neural network
modules.
Future work includes using this framework in
differentiable forms of topological transforms
(as in \cite{roell2023differentiable}) and incorporating
these transforms into neural network architectures.
Additionally, we believe that this work will encourage further investigation
into vectorized methods for computing other topological functions and
transforms.

\section*{Acknowledgements}
We thank Jack Ruder and Jacob Sriraman for their valuable assistance in
developing and refining portions of the methodology and code used
in this work.
Computational efforts were performed on the Tempest High Performance
Computing System, operated and supported by the University Information
Technology Research Cyberinfrastructure (RRID:SCR\_026229)
at Montana State University.

\bibliography{main}

\appendix
\section{Examples of Tensor Operations}
\label{append:tensor-examples}

In this appendix, we provide a few examples to illustrate the tensor operations
defined in \ssref{tensorops}.
Consider the following~tensors:
\[
    T = \begin{bmatrix}
        1 & 2 \\
        3 & 4 \\
        5 & 6
    \end{bmatrix}, \quad
    U = \begin{bmatrix}
        1 \\
        2
    \end{bmatrix}, \quad
    V = \begin{bmatrix}
        5
    \end{bmatrix}
\]

The $\cumsum$ operation computes the cumulative sum of each row of the tensor.
For the tensor~$T$, we see
\[
    \cumsum(T) = \begin{bmatrix}
        1 & 1 + 2 \\
        3 & 3 + 4 \\
        5 & 5 + 6
    \end{bmatrix} = \begin{bmatrix}
        1 & 3 \\
        3 & 7 \\
        5 & 11
    \end{bmatrix}.
\]

The $\rmax$ operation computes the maximum value with respect to a specified axis.
That is, for the tensor $T$ and axis $p = 1$, we have
\[
    \rmax(T, 1) = \begin{bmatrix}
        \max(1, 2) \\
        \max(3, 4) \\
        \max(5, 6)
    \end{bmatrix} = \begin{bmatrix}
        2 \\
        4 \\
        6
    \end{bmatrix}.
\]

Next, we consider the advanced indexing operation, using $U$ to index into $T$.
Note that~$T$ has shape~$(3, 2)$ and $U$ has shape $(2, 1)$, so the resulting tensor
has shape $(2, 1, 2)$.
Then, for each $k \in [[2]]$ we have
\[
    T[U] = \begin{bmatrix}
            T[U[0, 0], k] \\
            T[U[0, 1], k]
        \end{bmatrix} =
        \begin{bmatrix}
            T[1, k] \\
            T[2, k]
        \end{bmatrix} =
        \begin{bmatrix}
            \begin{bmatrix}
                3 & 4
            \end{bmatrix} \\
            \begin{bmatrix}
                5 & 6
            \end{bmatrix}
        \end{bmatrix}.
\]

To demonstrate the $\scadd$ operation, we need to ensure dimensions align correctly, so consider the transpose $S = T^T$, which is
\[
    S = \begin{bmatrix}
        1 & 3 & 5 \\
        2 & 4 & 6
    \end{bmatrix},
\]
Next, we compute $\scadd(U, V)(S)$.
To do so, we compute the difference tensor $D$, which is used to modify $S$ in place.
The tensor $D$ has the same shape as $S$, which is $(2, 3)$.
Because~$V$ has shape $(1)$, we note that $k=0$ is the only possible value of
$k$ in the definition of $D$ given in \eqref{difftensor}.  As a result, we look
for indices $i,j$ such that $U[i,0]=j$.
Thus, we start with $D=\zeros((2,3))$ and compute the values of $D$ by adding
the following values:
\begin{enumerate}
    \item For the first row ($i = 0$), the index from $U$ is $U[0,0] = 1$,
        and the value from $V$ is~$5$.
        Thus, we add the value $5$ to the element at position $(i, U[i, k]) =
        (0,U[0,0])= (0, 1)$,
        so the first row of $D$ is $[0, 5, 0]$.
    \item For the second row ($i = 1$), the index from $U$ is $U[1, 0] = 2$,
        and the value from $V$ is~$5$.
        Thus, we add the value $5$ to the element at position $(i, U[i, k]) =
        (1,U[1,0]) = (1, 2)$,
        so the second row of $D$ is~$[0, 0, 5]$.
\end{enumerate}
Thus, the difference tensor is
\[
    D = \begin{bmatrix}
        0 & 5 & 0 \\
        0 & 0 & 5
    \end{bmatrix},
\]
and we have
\[
    \scadd(U, V)(S) = S + D = \begin{bmatrix}
        1 + 0 & 3 + 5 & 5 + 0 \\
        2 + 0 & 4 + 0 & 6 + 5
    \end{bmatrix} = \begin{bmatrix}
        1 & 8 & 5 \\
        2 & 4 & 11
    \end{bmatrix}.
\]
Recall that the $\scadd$ operation modifies the input tensor in place,
so the new value of $S$ is
\[
    S = \begin{bmatrix}
        1 & 8 & 5 \\
        2 & 4 & 11
    \end{bmatrix}.
\]

\end{document}